\documentclass[prl,showpacs,twocolumn,amsmath]{revtex4}

\def\be{\begin{equation}}
\def\ee{\end{equation}}

\newcommand \dd{\mathrm{d}}

\newcommand \e{\text{e}}

\usepackage{graphicx,epsfig,latexsym,makeidx,amsmath}
\usepackage{color}
\usepackage{float}
\usepackage{subcaption} % 使用 subfigure 环境
\usepackage{amssymb} % 数学符号支持
\usepackage{caption}    % 增强 caption 功能
\usepackage[colorlinks,linkcolor=blue,anchorcolor=blue,citecolor=green,urlcolor=blue]{hyperref}
\usepackage[english]{babel}
\usepackage{booktabs}  % 增强表格线的美观性（可选）
\usepackage{array}     % 允许自定义列格式
\usepackage{caption}   % 控制表格标题格式
\usepackage{diagbox}    % 生成带斜线的单元格
\usepackage{multirow}
\usepackage[a4paper,top=2cm,bottom=2cm,left=1.7cm,right=1.7cm,marginparwidth=1.75cm]{geometry}
\usepackage{amsthm}

\begin{document}
% Useful packages

%%%%%%%%%%%%%%%%%%%%%%%%%%%%%%%%%%%%%%%%%%%%%%%%%%%%%%%%%%%%%%%%%%%%%%%%%%%%%%%%%%%%%%

	\newcommand \tr{\textcolor{red}}
	\newcommand{\PR}[1]{\ensuremath{\left[#1\right]}} % parenteses rectos do tamanho adequado
	\newcommand{\PC}[1]{\ensuremath{\left(#1\right)}} % parenteses curvos do tamanho adequado
	\newcommand{\PX}[1]{\ensuremath{\left\lbrace#1\right\rbrace}} % chavetas do tamanho adequado
	\newcommand{\BR}[1]{\ensuremath{\left\langle#1\right\vert}} % Bra do tamanho adequado
	\newcommand{\KT}[1]{\ensuremath{\left\vert#1\right\rangle}} % Ket do tamanho adequado
	\newcommand{\MD}[1]{\ensuremath{\left\vert#1\right\vert}} % modulo do tamanho adequado
	%%%%%%%%%%%%%%%%%%%%%%%%%%%%%%%%%%%%%%%%%%%%%%%%%%%%%%%%%%%%%%%%%%%%%%%%%%%%%%%%%%%

	\title{Quasinormal modes of thick branes in $f(R)$ gravity}

	\author{Yu-Peng E$^{a}$ $^{b}$, Chun-Chun Zhu$^{c}$, and Yu-Xiao Liu$^{a}$ $^{b}$ \footnote{liuyx@lzu.edu.cn, corresponding author}}
	
	\affiliation{$^{a}$Lanzhou Center for Theoretical Physics, Key Lacboratory for Quantum Theory and Applications of the MoE, Key Laboratory of Theoretical Physics of Gansu Province,  Gansu 
		Provincial Research Center for Basic Disciplines of Quantum Physics, Lanzhou University, Lanzhou 730000, China\\
		$^{b}$Institute of Theoretical Physics and Research Center of Gravitation, School of Physical Science and Technology,  Lanzhou University, Lanzhou 730000, China\\
		$^{c}$School of Mathematics and Physics, Jingchu University of Technology, Jingmen 448000, China\\}
	\begin{abstract}
		We systematically investigate the quasinormal modes of thick branes in $f(R)$ gravity by numerically solving the Schr\"odinger-like perturbation equation of gravitational 
		perturbations. To ensure the reliability  of the results, we employ three complementary methods: the asymptotic iteration method, the direct integration of the wave equation, and the time-domain numerical evolution. We analyze how the model parameters influence the shape of the effective potential of gravitational perturbations and find that the structure of the potential barrier plays a significant role in shaping the quasinormal frequency spectrum. The results obtained from the three methods exhibit strong consistency, thereby ensuring the reliability of the calculations. In particular, the real parts of the quasinormal frequencies exhibit an approximately arithmetic progression, suggesting that the quasi-localized states can be understood as resonances between the barriers.
	\end{abstract}
	
	\maketitle
	\flushbottom

	\section{Introduction}
	\label{sec:introduction}
	To date, the LIGO–Virgo–KAGRA (LVK) collaboration has reported more than 300 gravitational wave (GW) events~\cite{LIGOScientific:2018mvr,LIGOScientific:2020ibl,KAGRA:2021vkt,LIGOScientific:2025slb}, opening a new observational window for probing fundamental physics and the structure of the Universe. The steadily increasing number of GW detections has enabled stringent tests of Einstein’s general relativity (GR) in the strong-field regime~\cite{LIGOScientific:2016lio,LIGOScientific:2018dkp,LIGOScientific:2019fpa,LIGOScientific:2020tif,LIGOScientific:2021sio,LIGOScientific:2025rid,LIGOScientific:2025obp}. Among all reported events, the recently observed GW250114 stands out as the most powerful GW signal detected so far~\cite{LIGOScientific:2025rid,LIGOScientific:2025obp}. Its exceptionally high network signal-to-noise ratio has allowed, for the first time, a clear identification of the quasinormal modes (QNMs) signal during the ring-down phase, enabling a high-precision analysis of its spectral properties.
	
	In gravitational physics, QNMs are complex-frequency solutions to perturbation equations that characterize the ring-down phase of a system as it returns to equilibrium. These modes are defined by discrete complex frequencies, where the real part corresponds to the oscillation frequency and the imaginary part reflects the damping rate due to energy dissipation. Although QNMs are most prominently studied in the context of black holes~\cite{Kokkotas:1999bd,Berti:2009kk,Vishveshwara:1970zz,Nollert:1996rf,Onozawa:1995vu,Andersson:1996xw,Onozawa:1996ux,Leung:1997was,Zhang:2006hh,Miranda:2008vb}, where they capture the relaxation dynamics of a perturbed black hole and are connected to gravitational-wave observables, their conceptual relevance extends to other physical systems~\cite{Horowitz:1999jd,Berti:2009kk,Tan:2024url,Konoplya:2011qq,Konoplya:2003dd}.
	
	The study of resonant or dissipative modes, often referred to as QNMs in non-Hermitian systems, plays a ubiquitous role across multiple branches of wave-dominated physics, including continuum mechanics, acoustics, electrodynamics, and quantum theory~\cite{Demesy:23,PhysRevA.49.3057,PhysRevX.11.041020}. This broad applicability implies that mathematical techniques originally developed for black hole QNMs can be fruitfully adapted to analyze the stability and dynamic response of other gravitational configurations, such as braneworlds~\cite{Kanti:2001cj,Seahra:2005wk,Cardoso:2001bb,Clarkson:2005mg,Toshmatov:2016bsb,Jia:2024sdk,Jia:2024pdk,Tan:2024qij,Bohra:2023vls,Lemos:2022vhw}.
	
	Braneworld models propose that our observable universe is a $(3+1)$-dimensional membrane, or ``braneworld" embedded within a higher-dimensional ``bulk" spacetime. A central objective of these models is to reproduce an effective four-dimensional gravitational theory on the brane. Over the past two decades, braneworld scenarios have attracted considerable interest as promising frameworks for addressing the gauge hierarchy problem in particle physics~\cite{5,Randall:1999vf}. Notable among these are the warped extra-dimensional models proposed by Randall and Sundrum—specifically the RS-1 and RS-2 models~\cite{5,Randall:1999vf}. In the 
	RS-1 scenario, the extra dimension is compactified and finite, thereby preserving four-dimensional Newtonian gravity at large scales. In contrast, the RS-2 model features an infinite extra dimension, yet still recovers the Newtonian gravitational potential on the brane, owing to 
	the warped geometry of the bulk. Both RS-1 and RS-2 represent idealized thin-brane configurations with zero thickness. By integrating concepts from domain wall models in flat spacetime~\cite{6}, the RS-2 framework has been generalized to thick-brane scenarios~\cite{7,8,9}. %The literature \cite{10,11,12} provides a detailed description of the braneworld theory.
	
	Within the framework of General Relativity, thick brane models offer a compelling refinement over their thin-brane counterparts by representing our four-dimensional universe as a dynamic domain wall or a solitonic structure with finite width (or ``thickness") embedded in a higher-dimensional bulk. This approach, often realized by coupling gravity to one or more scalar fields~\cite{7,9,Campos:2001pr,Bazeia:2003aw,Gremm:1999pj,DeWolfe:1999cp,Kehagias:2000au,Kobayashi:2001jd,Dzhunushaliev:2009va,Liu:2007ku,Bazeia:2014poa,Guo:2011wr}, naturally resolves the singularities associated with infinitely thin branes and provides a smooth localization mechanism for various matter and gravitational fields. The geometry is typically warped, as in the classic Randall-Sundrum scenario, but with a smooth warp factor driven by the scalar field dynamics, leading to the successful recovery of effective four-dimensional gravity~\cite{Rubakov:1983bb,Kehagias:2000au,Liu:2012rc,DeWolfe:1999cp,Gremm:1999pj,Kobayashi:2001jd,Liu:2007gk,Dzhunushaliev:2009va,Guo:2011wr,Bazeia:2014poa}.
	
	The exploration of brane worlds has been significantly extended beyond General Relativity to various modified gravity theories, which introduce higher-order curvature invariants or non-minimal couplings~\cite{Balcerzak:2010kr,Bazeia:2014poa,Maeda:2003vq,Liu:2012rc}. Among these, $f(R)$ gravity has been a particularly active area of research. In this framework, the standard Einstein-Hilbert action is generalized to an arbitrary function of the Ricci scalar $R$. Studies of $f(R)$ thick brane models~\cite{Bazeia:2014poa,Zhong:2015pta,38,37} revealed a richer variety of warp factors and scalar field configurations, demonstrating that the higher-order curvature terms can profoundly influence the brane's internal structure, stability, and the spectrum of gravitational modes. These models not only maintain the desirable features of standard thick branes but also exhibit new phenomena, such as changes in the localization properties of fermions and the potential for novel gravitational resonances~\cite{Liu:2009mga,Zhang:2016ksq}. Comprehensive reviews of related research of thick branes can be found in Refs.~\cite{10,11,12}.
	
	Recent studies have shown that, in addition to the zero mode localized on the brane, there exists a discrete set of QNMs~\cite{Jia:2024sdk,Jia:2024pdk,Tan:2023cra}, within the seemingly continuous spectrum of massive Kaluza-Klein (KK) modes, whose spectral structure is closely tied to the geometry and internal structure of the brane. In Ref.~\cite{Tan:2023cra}, the QNMs of a free scalar field in the brane-world scenario were obtained for the first time within the framework of General Relativity, revealing a close relationship between QNMs and resonant states. In Ref.~\cite{Zhu:2024gvl}, it was shown that long-lived QNMs can give rise to gravitational echo phenomena. Ref.~\cite{Jia:2024sdk} further investigated how different types of branes affect the quasinormal spectra, demonstrating that the brane structure plays a crucial role in determining the quasinormal frequencies (QNFs). Moreover, QNMs have also been found to exist in brane-world models based on Rastall gravity~\cite{Tan:2024url}, suggesting that such oscillatory behaviors are a generic feature beyond Einstein’s theory.
	In this work, we extend these studies by exploring the QNMs of a flat brane in the framework of $f(R)$ gravity, aiming to understand how higher-order curvature corrections influence the spectrum and stability of gravitational perturbations.
		
	The structure of this paper is organized as follows. In Sec.~\ref{sec:method}, we review the thick brane solutions in $f(R)$ gravity and present the formalism for linear gravitational perturbations. In Sec.~\ref{sec:main_section}, we compute the QNFs using three independent methods: the asymptotic iteration method (AIM), the numerical evolution, and the direct integration. We then analyze how the 
	QNMs depend on the model parameter $\alpha$ in the specific case of $f(R)=R+\alpha R^2$. 
	Finally, in~\ref{sec:conclusion}, we summarize the main findings and present our conclusions.
	
	\section{Thick brane model in $f(R)$ gravity}
	\label{sec:method}
	In this section, we review the five-dimensional thick brane model in  $ f(R) $ gravity. The action is given by
	\begin{equation}
		\displaystyle
		S = \int d^5x \sqrt{-g} \left( \frac{1}{2\kappa_5^2}  f(R)   - \frac{1}{2} g^{MN} \partial_M \varphi \partial_N \varphi - V(\varphi) \right),
	\end{equation}
	where $ f(R) $ is a function of the scalar curvature $R$, and $ V(\varphi) $ is the potential of the scalar field $\varphi$. In this paper, we set $ \kappa_5 = 1 $. The field equations are obtained by varying the action with respect to the metric $g_{MN}$ and the scalar field $\varphi$:
	\begin{equation}
		\displaystyle
		f_R R_{MN} - \frac{1}{2} f g_{MN} - \left( \nabla_M \nabla_N - g_{MN}\Box^{(5)} \right) f_R = T_{MN}, 
		\label{2z}
	\end{equation}
	\begin{equation}
		\displaystyle
		\Box^{(5)} \varphi \equiv g^{MN} \nabla_M \nabla_N \varphi = V_{\varphi}, 
		\label{3z}
	\end{equation}
	where $ f_R \equiv \frac{df(R)}{dR} $, $V_\varphi \equiv \frac{dV(\varphi)}{d\varphi}$, and $T_{MN} = \partial_M \varphi \partial_N \varphi - g_{MN} \left( \frac{1}{2} g^{AB} \partial_A \varphi \partial_B \varphi + V(\varphi) \right)$ is the energy-momentum tensor. The metric describing a static flat brane is given by
	\begin{equation}
		\displaystyle
		ds^2 = g_{MN} dx^M dx^N = e^{2A(y)} \eta_{\mu\nu} dx^\mu dx^\nu + dy^2,
		\label{metric}
	\end{equation}
	where $ e^{A(y)} $ is the warp factor. After performing the coordinate transformation $dy=e^A dz$, the metric can be recast in a conformally flat form: 
	\begin{equation}
		\displaystyle
		ds^2 = e^{2A(z)}( \eta_{\mu\nu} dx^\mu dx^\nu + dz^2). 
	\end{equation}
	Substituting the metric~\eqref{metric} into Eqs.~\eqref{2z} and \eqref{3z}, we obtain the 
	following explicit field equations
	\begin{equation}
		\displaystyle
		f+2f_R(4A^{\prime 2}+A^{\prime \prime})-6f^{\prime}_RA^{\prime}-2f^{\prime \prime}_R=\varphi^{\prime 2}+2V,
	\end{equation}
	\begin{equation}
		\displaystyle
		-8f_R(A^{\prime \prime}+A^{\prime 2})+8f^{\prime}_RA^{\prime}-f=\varphi^{\prime 2}-2V,
	\end{equation}
	\begin{equation}
		\displaystyle
		4A^{\prime}\varphi^{\prime}+\varphi^{\prime \prime}=V_{\varphi}.
	\end{equation}
	It can be shown that although there are three field equations, only two are independent. Now we have four unknown functions but only two independent equations. Thus, we should give two of them. In this paper, we consider $ f(R) = R + \alpha R^2 $. Exact analytical solutions for this model can be found in Ref.~\cite{36,Xu:2014jda,Bazeia:2013uva}.
	
	Next, we consider tensor perturbations of the metric in coordinates of $(x^{\mu},z)$:
	\begin{equation}
		\displaystyle
		ds^2 = e^{2A(z)} \left( (\eta_{\mu\nu} + h_{\mu\nu}) dx^\mu dx^\nu + dz^2 \right),
		\label{9}
	\end{equation}
	where $ h_{\mu\nu} $ is a transverse traceless tensor:
	\begin{equation}
		\displaystyle
	     \partial^\mu h_{\mu\nu} = 0, 
		~\eta^{\mu\nu} h_{\mu\nu} = 0.
		\label{10}
	\end{equation}
	The tensor perturbation can be decomposed as follows:
	\begin{equation}
		h_{\mu\nu}(x^i,t,z) = \left( a^{-\frac{3}{2}} f_R^{-\frac{1}{2}} \right) \epsilon_{\mu\nu}(x^i) H(t,z),
		\label{11}
	\end{equation}
	where $ a = e^{A} $, $\delta^{ij} \partial_i \partial_j \epsilon_{\mu\nu} = -p^2 \epsilon_{\mu\nu}$ with $p^2=\delta_{ij} p^i p^j$ the square of the three-dimensional momentum of the gravitational Kaluza-Klein mode. By substituting Eqs.~\eqref{9} and \eqref{11} into Eq.~\eqref{2z}, we obtain the evolution equation for the extra-dimensional component:
	\begin{equation}
		\partial_t^2 H(t, z) - \partial_z^2 H(t, z) +W(z) H(t, z) = -p^2 H(t, z).
        \label{waveeq}
	\end{equation}
	Here, the effective potential $W(z)$ is~\cite{38}
	\begin{equation}
		\displaystyle
		W(z) = \frac{3}{4} \frac{(\partial_z a)^2}{a^2} + \frac{3}{2} \frac{\partial_z^2 a}{a} + \frac{3}{2} \frac{\partial_z a \partial_z f_R}{a f_R} - \frac{1}{4} \frac{(\partial_z f_R)^2}{f_R^2} + \frac{1}{2} \frac{\partial_z^2 f_R}{f_R}.
        \label{effp}
	\end{equation}
	Note that there are no explicit scalar-field terms appearing in Eqs.~\eqref{waveeq} and~\eqref{effp}. The reason is that, within the framework considered in this paper, the scalar field is minimally coupled to gravity, and we only focus on transverse-traceless tensor perturbations. More specifically, the scalar field affects the system only through the background equations, and its dynamical degrees of freedom do not directly couple to the TT tensor perturbations. As a result, when deriving the evolution equation for the extra-dimensional component from Eq.~\eqref{waveeq}, the scalar field does not appear as an explicit source term.
    
    Assuming $H(t, z) = \psi(z) e^{i \omega t}$, this equation reduces to the Schr\"odinger-like form \cite{38}:
	\begin{equation}
		\left(-\partial^2_z+ W(z)\right)\psi(z) = m^2 \psi(z),
		\label{13}
	\end{equation}
	where $ m^2 = \omega^2 - p^2$ is the four-dimensional effective mass of the Kaluza-Klein mode. 
	The Schrödinger-like equation \eqref{13} can be rewritten as
	\begin{equation}
		\displaystyle
		\mathcal{K} \mathcal{K}^\dagger \psi(z) = m^2 \psi(z),
	\end{equation}
	where the two operators are given by
	\begin{align}
		\mathcal{K} & =\partial_z + \frac{3}{2} \frac{\partial_z a}{a} + \frac{1}{2} \frac{\partial_z f_R}{f_R}, \\
		\mathcal{K}^\dagger & =-\partial_z + \frac{3}{2} \frac{\partial_z a}{a} + \frac{1}{2} \frac{\partial_z f_R}{f_R}.
	\end{align}
	By setting $m=0$, we obtain the gravitational zero mode:
	\begin{equation}
		\displaystyle
		\psi^{(0)}(z) = N_0 a^{3/2}(z) f_R^{1/2}(z).
	\end{equation}
	It is easy to show that the gravitational zero mode for the $f(R)$-brane found in Ref.~\cite{36}  satisfies
	\begin{equation}
		\displaystyle
		\int_{-\infty}^{\infty} |\psi^{(0)}(z)|^2 \, dz < \infty.
	\end{equation}
	This result indicates that the zero mode is normalizable and hence is localized on the brane. Particularly, if $ f_R(z) = 1 + 2\alpha R(z) = 0 $ at $ z = \pm z_0 $, the effective potential $W(z)$ is divergent at these points, which will result in special localization of 
	graviton zero mode~\cite{Xu:2014jda}.
	
	Next, we will consider two types of warp factor solutions and calculate the corresponding QNMs under tensor perturbations in the next section.
	
	\subsection{Model A}
	In the first model, we consider $f(R)=R + \alpha R^2$ and $a(y) = \text{sech}^B (ky)$ with $B$ a positive integer~\cite{Xu:2014jda}. The equation for the scalar field is  
	\begin{eqnarray}
		\varphi'^2(y) &= B k^2\text{sech}^2(ky)\big(
		\frac{3}{2} - 4 \alpha k^{2}( 5B^{2} + 16B + 8 \notag \\
		& - (5B^{2} + 32B + 12)\text{sech}^{2}(ky))
		\big). \label{phiprime}
	\end{eqnarray}
	Note that the parameter $\alpha$ used here corresponds to $-\alpha$ in Ref.~\cite{Bazeia:2013uva}. Therefore, the condition $\varphi'^2 \ge 0$ requires~\cite{Xu:2014jda}
	\begin{eqnarray}
		\alpha_1 \equiv -\frac{3}{32(1+4B)k^2} \le \alpha \le \frac{3}{8(8+16B+5B^2)k^2} \equiv \alpha_2.
	\end{eqnarray}

    	\begin{figure}[htbp]  % 图片浮动体环境
		\centering        % 图片居中
		\includegraphics[width=0.27\textwidth]{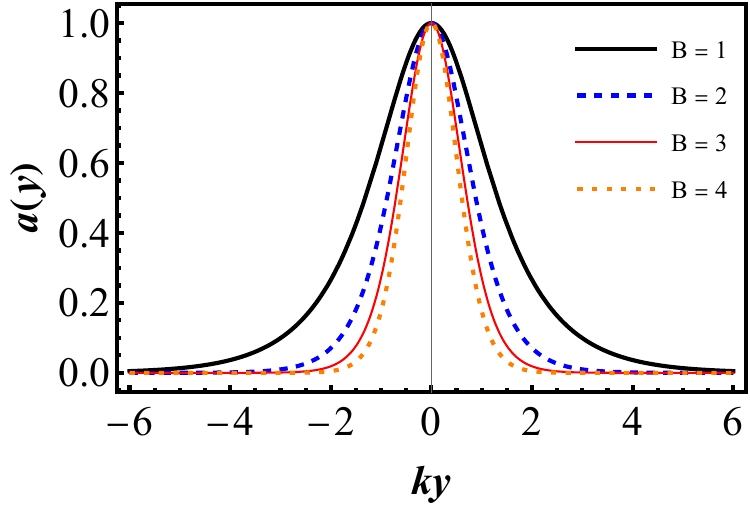}  % 图片文件名 (不包含扩展名) 和尺寸
		\caption{ The warp factor in Model A with different values of the parameter $B$.}  % 图片说明
		\label{1(a)}  % 图片标签，方便引用
	       \end{figure}
    	\begin{figure}[ht]
		\centering
		\begin{subfigure}{0.2\textwidth} % 设置子图的宽度
			\centering
			\includegraphics[width=\linewidth]{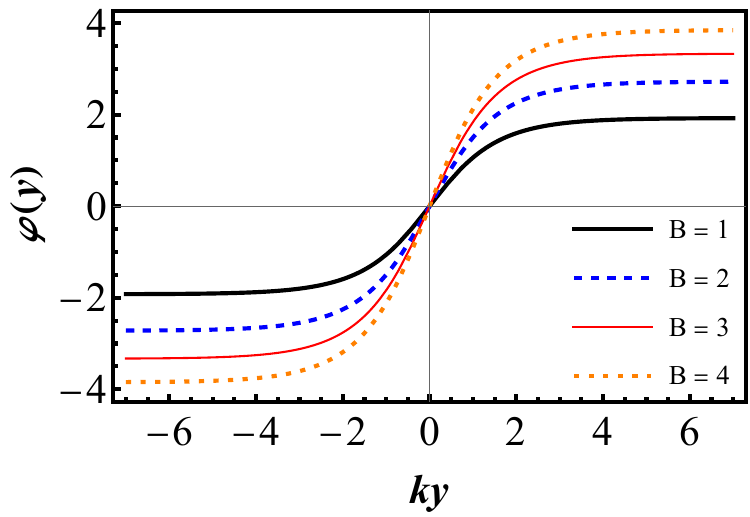} % 替换为你的图片文件名
			\caption{$\alpha=0$}
			\label{1(b)}
		\end{subfigure}
		\hspace{0.025\textwidth} % 添加水平间距
		\begin{subfigure}{0.2\textwidth} % 设置子图的宽度
			\centering
			\includegraphics[width=\linewidth]{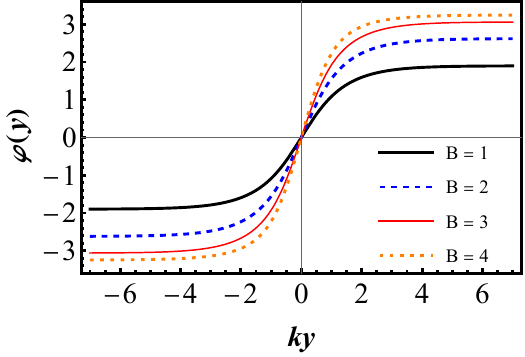} % 替换为你的图片文件名
			\caption{$\alpha=0.002$}
			\label{1(c)}
		\end{subfigure}
		\vspace{0.5cm} % 两行图片之间的垂直间距
		\begin{subfigure}{0.2\textwidth} % 设置子图的宽度
			\centering
			\includegraphics[width=\linewidth]{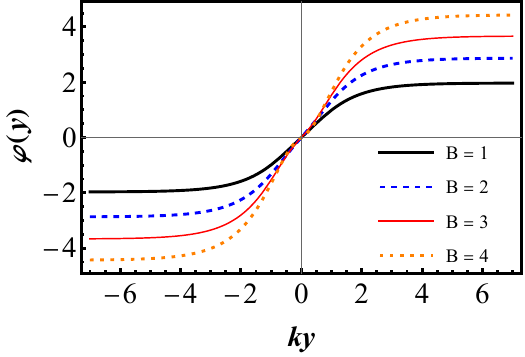} % 替换为你的图片文件名
			\caption{$\alpha=-0.005$}
			\label{1(d)}
		\end{subfigure}
		\hspace{0.025\textwidth} % 添加水平间距
		\begin{subfigure}{0.2\textwidth} % 设置子图的宽度
			\centering
			\includegraphics[width=\linewidth]{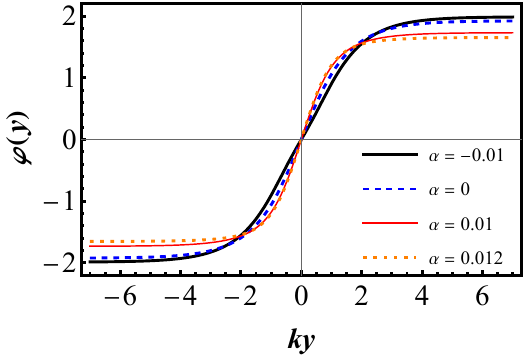} % 替换为你的图片文件名
			\caption{$B=1$}
			\label{1(e)}
		\end{subfigure}
		\vspace{0.5cm} % 两行图片之间的垂直间距
		\begin{subfigure}{0.2\textwidth} % 设置子图的宽度
			\centering
			\includegraphics[width=\linewidth]{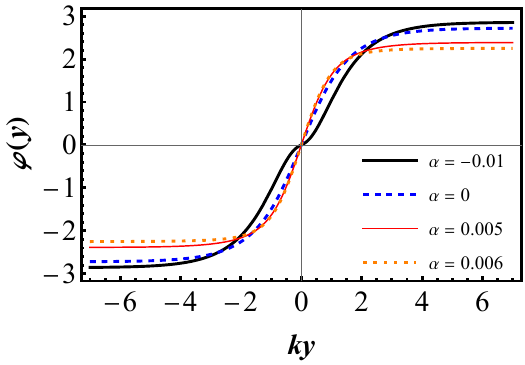} % 替换为你的图片文件名
			\caption{$B=2$}
			\label{1(f)}
		\end{subfigure}
		\hspace{0.025\textwidth} % 添加水平间距
		\begin{subfigure}{0.2\textwidth} % 设置子图的宽度
			\centering
			\includegraphics[width=\linewidth]{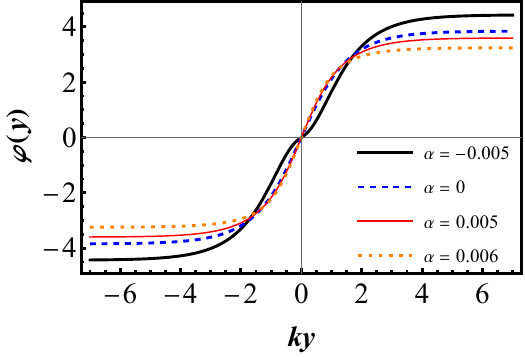} % 替换为你的图片文件名
			\caption{$B=4$}
			\label{1(g)}
		\end{subfigure}
		\caption{The scalar field $\varphi(y)$ in Model A with different values of the parameters $B$ and $\alpha$.
		}
		\label{fig1:two_images}
	\end{figure}

    \begin{figure}[htbp]
		\centering
		\begin{subfigure}{0.2\textwidth} % 设置子图的宽度
			\centering
			\includegraphics[width=\linewidth]{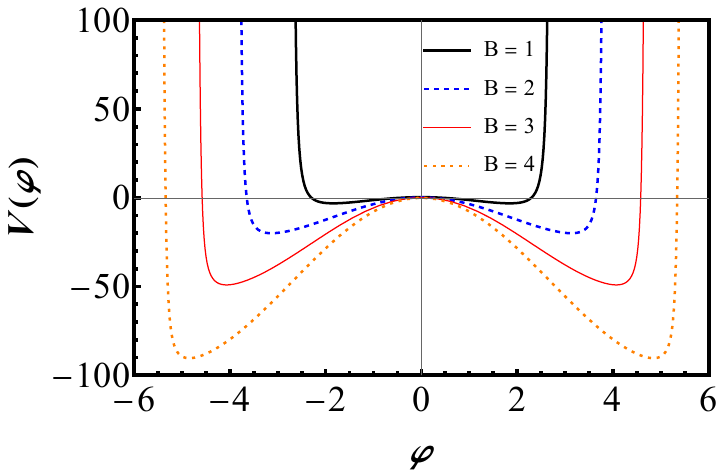} % 替换为你的图片文件名
			\caption{$\alpha=0$}
			\label{33(A)}
		\end{subfigure}
		\hspace{0.025\textwidth} % 添加水平间距
		\begin{subfigure}{0.2\textwidth} % 设置子图的宽度
			\centering
			\includegraphics[width=\linewidth]{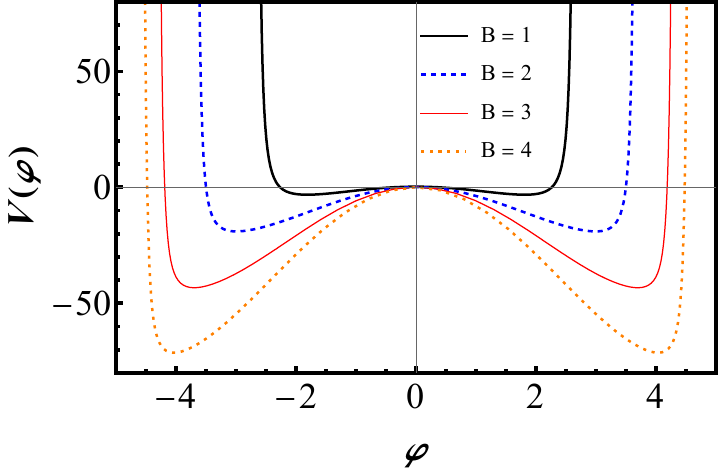} % 替换为你的图片文件名
			\caption{$\alpha=0.002$}
			\label{33(B)}
		\end{subfigure}
		\vspace{0.5cm} % 两行图片之间的垂直间距
		\begin{subfigure}{0.2\textwidth} % 设置子图的宽度
			\centering
			\includegraphics[width=\linewidth]{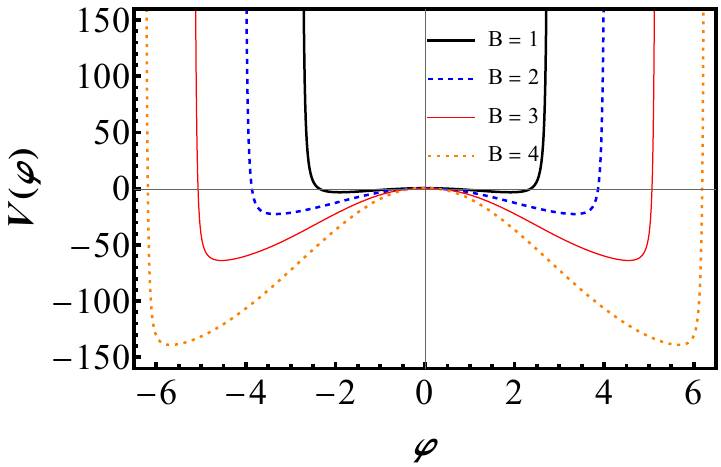} % 替换为你的图片文件名
			\caption{$\alpha=-0.005$}
			\label{33(C)}
		\end{subfigure}
		\hspace{0.025\textwidth} % 添加水平间距
		\begin{subfigure}{0.2\textwidth} % 设置子图的宽度
			\centering
			\includegraphics[width=\linewidth]{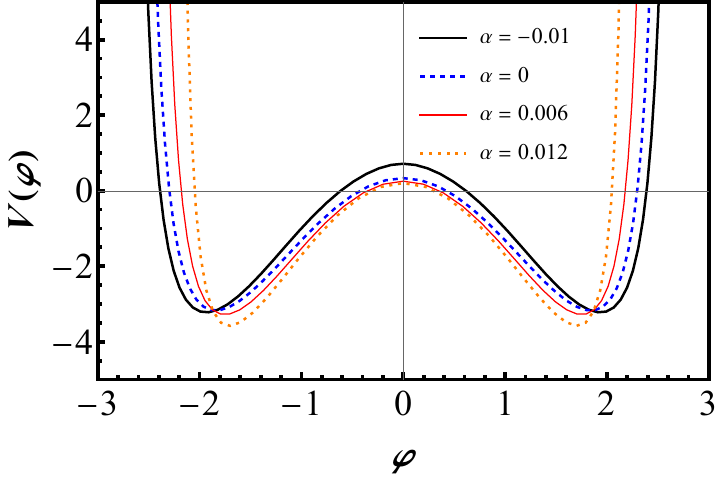} % 替换为你的图片文件名
			\caption{$B=1$}
			\label{33(D)}
		\end{subfigure}
		\vspace{0.5cm} % 两行图片之间的垂直间距
		\begin{subfigure}{0.2\textwidth} % 设置子图的宽度
			\centering
			\includegraphics[width=\linewidth]{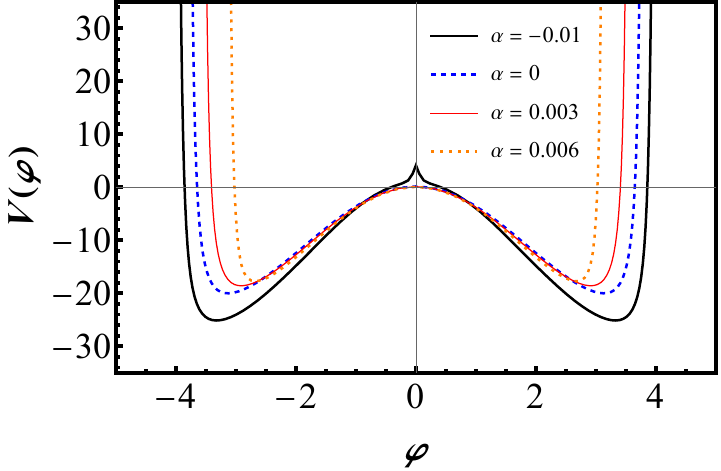} % 替换为你的图片文件名
			\caption{$B=2$}
			\label{33(E)}
		\end{subfigure}
		\hspace{0.025\textwidth} % 添加水平间距
		\begin{subfigure}{0.2\textwidth} % 设置子图的宽度
			\centering
			\includegraphics[width=\linewidth]{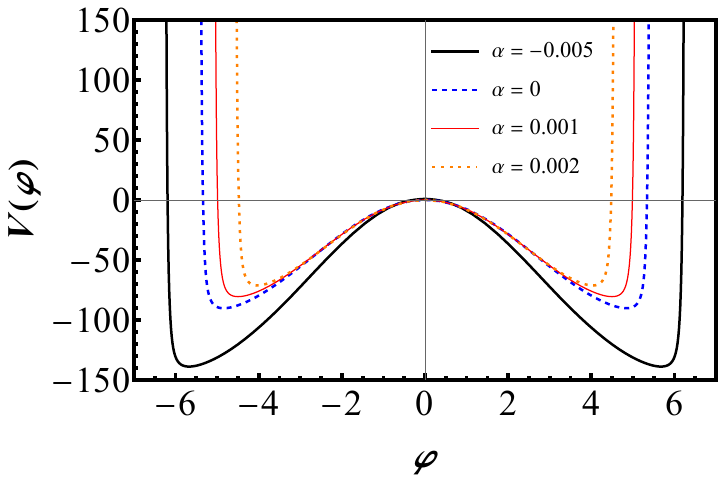} % 替换为你的图片文件名
			\caption{$B=4$}
			\label{33(F)}
		\end{subfigure}
		\caption{The scalar potential $V(\varphi)$ in Model A with different values of the parameters $B$ and $\alpha$.
		}
		\label{fig3:6_images}
	\end{figure}
	In addition, considering the $ \mathbb{Z}_2 $ symmetry of the background, the scalar field is chosen to be a kink configuration satisfying $ \varphi(0)=0 $. Figure~\ref{1(a)} shows the shape of the warp factor, while Fig.~\ref{fig1:two_images} shows the numerical solutions of the scalar field with different values of $ B $ and $\alpha$. As shown in Figs.~\ref{1(b)},~\ref{1(c)}, and \ref{1(d)}, the scalar field with fixed $\alpha$ exhibits a single-kink profile and its vacuum expectation value increases with $B$.  For the case of fixed $B$ (see Figs.~\ref{1(e)}, ~\ref{1(f)}, and \ref{1(g)}), the vacuum expectation value of the scalar field decreases with $\alpha$. 
    
    Correspondingly, two minima of the scalar potential with fixed $\alpha$ decrease with $B$, as shown in Figs.~\ref{33(A)},~\ref{33(B)}, and~\ref{33(C)}.  As shown in Figs.~\ref{33(D)},~\ref{33(E)}, and~\ref{33(F)}, for the case of fixed $B$, the minima of the scalar potential increase with $|\alpha|$ as $B=1$, and the minima of the scalar potential decrease with $\alpha$ as $B>1$. Accordingly, the value of the scalar potential at $\varphi=0$ decreases with $\alpha$ as $B\leq2$; otherwise, the change is not significant. Throughout this paper, we take $k=1$ in our calculations.
	
	Regarding the effective potential, when $ B > 1 $, there is no analytical solution in the $ z $-coordinate. But we can write its expression in the $y$-coordinate:
	\begin{widetext}
    \begin{align}
		W(z(y)) &= \frac{3}{4} \left(a'(y)\right)^2 
		+ \frac{2}{3} \left(a'^2 (y)+ a''(y)\right)
		+ \frac{3 \alpha a(y) a'(y) R'(y)}{1 + 2\alpha R(y)} \nonumber \\
		&\quad - \frac{\alpha^2 a^2(y) {R'^2(y)}}{(1 + 2\alpha R(y))^2}  
		+  \frac{\alpha (a(y)a'(y)R'(y) + a^2(y) R''(y))}{1 + 2\alpha R(y)} \nonumber \\
		&= \frac{B k^2 \text{sech}^{2 B}(k y)}
		{4 (1 + 16 B k^2 \alpha \text{sech}^2(k y) - 40 B^2 k^2 \alpha \tanh^2(k y))^2} \Bigg( 
		-512 B (2 +5 B) k^4 \alpha^2  \nonumber \\
		&\quad \text{sech}^6(k y)+ \frac{3}{8} (-4-5 B + 5 B \cosh(2 k y))
		\big(1 + 40 B^2 k^2 \alpha + (1 - 40 B^2 k^2 \alpha)  \nonumber \\
		&\quad \cosh^2 (2 k y)\big) \text{sech}^6(k y) - 8 k^2 \alpha (-4 - 30 B - 35 B^2 + (4 + 18 B + 35 B^2) \nonumber \\
		&\quad \cosh(2 k y)) \big(-1 - 40 B^2 k^2 \alpha + (-1 + 40 B^2 k^2 \alpha) \cosh(2 k y)\big) \text{sech}^6(k y) + 32 k^2 \nonumber \\
		&\quad  \alpha\text{sech}^4(k y) 
		\big(-2 - 5 B - 48 B^2 k^2 \alpha + 8 B (4 + 32 B + 55 B^2) k^2 \alpha \tanh^2(k y)\big) \Bigg).
	\end{align}
    \end{widetext}
	When $ B = 1 $ and $ \alpha = 0 $, the effective potential $ W $ has an analytical solution in the $ z $-coordinate, which is given by
	\begin{equation}
		W(z) = \frac{3 k^2\left( -9 + 5 \cosh \left( 2 \, \text{arcsinh}(kz) \right) \right)}{8 \left( 1 + k^2 z^2 \right)^2}.
		\label{23}
	\end{equation}

	The higher-order derivative terms in $f(R)$ gravity theory are closely associated with the emergence of ghost fields \cite{28,29,Nojiri:2010wj,Nojiri:2017ncd}. To avoid such ghost instabilities, the coefficients of the higher-order curvature terms must satisfy specific consistency conditions. For $ f(R) = R + \alpha R^2 $, the condition is $ f_R = 1+2\alpha R > 0 $ \cite{Zhong:2015pta}. In this model, the scalar curvature $ R $ is given by
	\begin{equation}
		R = -4 \left( 5 A'^2(y) + 2 A''(y) \right).
		\label{24}
	\end{equation}
    
	Substituting $ A(y) = B  \ln(\text{sech}(k y)) $ into  Eq.~\eqref{24}, we can get
	\begin{equation}
		R = -4 \left( -2 B k^2 \text{sech}^2(k y) + 5 B^2 k^2 \tanh^2(k y) \right).
	\end{equation}
	Accordingly, the scalar curvature $ R $ varies within the range $ -20 B^2 k^2 < R < 8 B k^2 $. Substituting this range of $R$ into the ghost-free condition $ f_R = 1 + 2\alpha R > 0 $, we derive the allowed range of the parameter $ \alpha$:
	\begin{align}
		-\frac{1}{16  k^2 B} < \alpha < \frac{1}{40  k^2 B^2}.
	\end{align}
	Based on previous work \cite{Xu:2014jda}, we consider the energy density of the scalar field:
	\begin{equation}
	\rho=\e^{2A(y) }\left(\frac{1}{2}\partial_M\varphi\partial_M\varphi + V (\varphi)\right).
\end{equation}
Solving $\frac{\dd^2\rho}{\dd y^2}|_{y=0}=0$ results in
	\begin{align}
		%k^2\alpha_1 & = -\frac{3}{32(1 + 4B)}, \\
		\alpha= -\frac{3 + 9B}{8k^2(16 + 60B + 49B^2)}\equiv\alpha_s.
		%k^2\alpha_2 & = \frac{3}{8(8 + 16B + 5B^2)}.
	\end{align}
	So $y = 0$ is an inflection point of $\rho$ when $\alpha=\alpha_s$, and the brane will have
	an internal structure when $\alpha\leqslant\alpha_s$.
	
\begin{figure}[ht]
		\centering
		% 第一行的两张图片
		\begin{subfigure}[b]{0.2\textwidth}
			\centering
			\includegraphics[width=\textwidth]{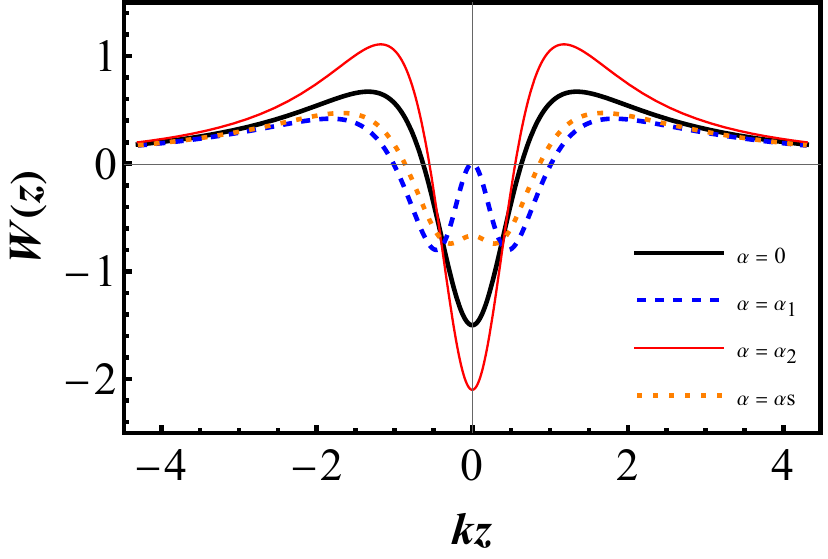} % 第一张图片
			\caption{ $ B = 1 $}
			\label{fig:img1}
		\end{subfigure}
		\hspace{0.025\textwidth} % 两张图片之间的水平间距
		\begin{subfigure}[b]{0.2\textwidth}
			\centering
			\includegraphics[width=\textwidth]{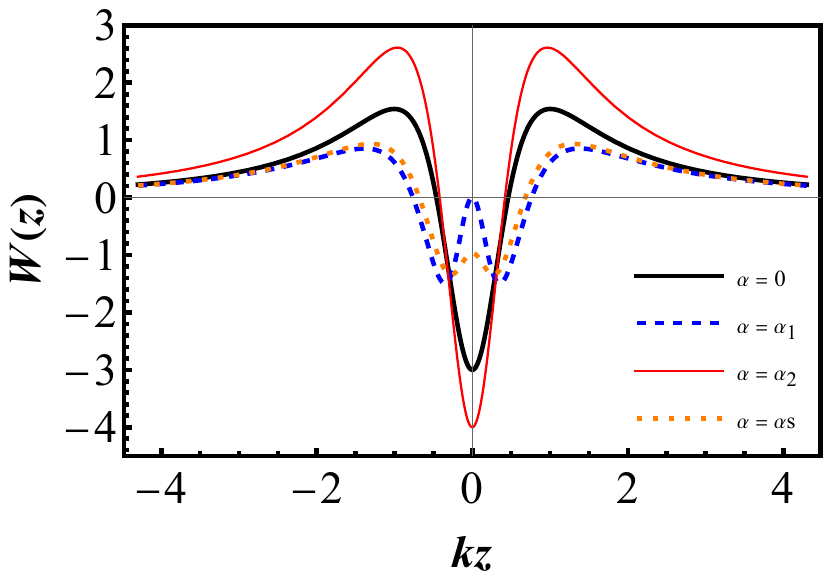} % 第二张图片
			\caption{ $ B = 2 $}
			\label{fig:img2}
		\end{subfigure}
		
		% 第二行的两张图片
		\vspace{0.5cm} % 两行图片之间的垂直间距
		\begin{subfigure}[b]{0.2\textwidth}
			\centering
			\includegraphics[width=\textwidth]{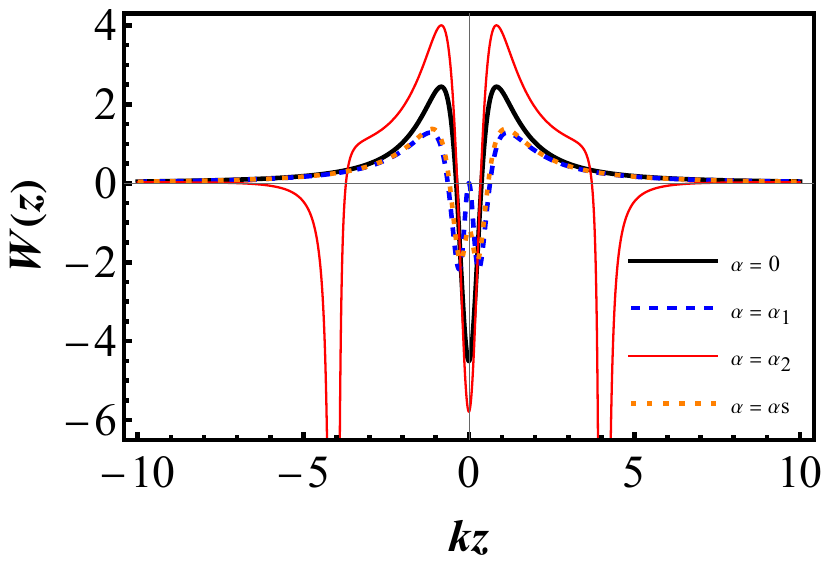} % 第三张图片
			\caption{ $ B = 3 $}
			\label{fig:img3}
		\end{subfigure}
		\hspace{0.025\textwidth} % 两张图片之间的水平间距
		\begin{subfigure}[b]{0.2\textwidth}
			\centering
			\includegraphics[width=\textwidth]{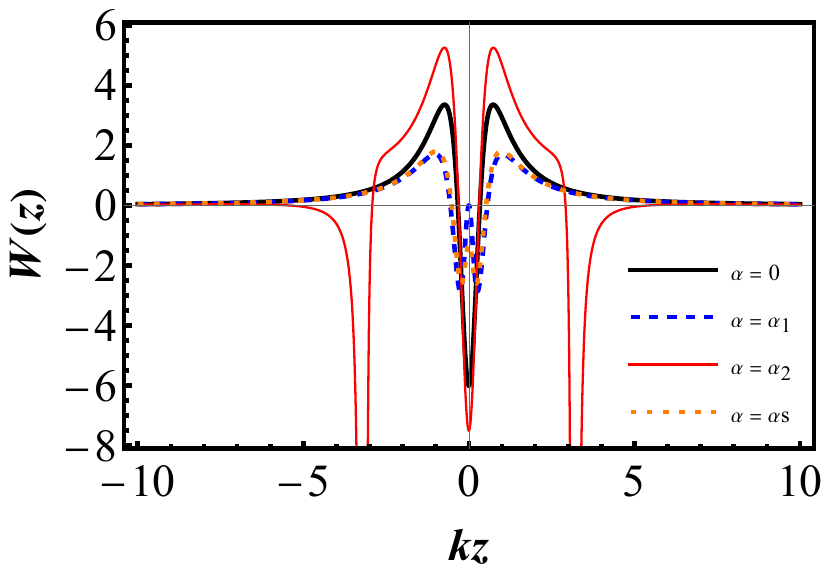} % 第四张图片
			\caption{ $ B = 4 $}
			\label{fig:img4}
		\end{subfigure}
		\caption{The shapes of the effective potential $ W(z) $ in Model A with different values of $B$ and $\alpha$.}
		\label{3}
	\end{figure}

	We present the effective potential $W(z)$  with different values of the parameters $\alpha$ and $B$ in Fig.~\ref{3}. The effective potential $W(z)$ is highly sensitive to variations in the parameter $\alpha$. For fixed $\alpha$, as $B$ increases, the potential barriers become higher, the potential well becomes deeper, and divergence points appear outside the barriers. For  fixed $B$, as $\alpha$ decreases, the barrier height becomes lower, a small peak gradually emerges at $z = 0$, and the divergence points outside the barriers disappear.
	
	\subsection{Model B}
	
\begin{figure}[ht]  % 图片浮动体环境
		\centering        % 图片居中
		\includegraphics[width=0.27\textwidth]{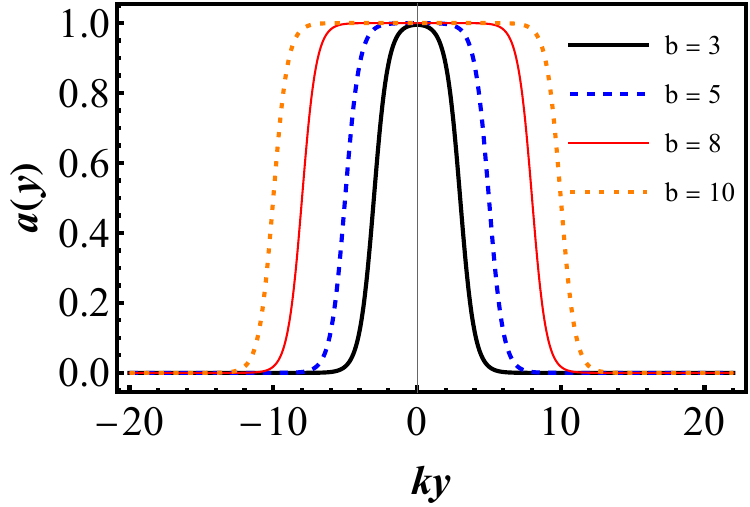}  % 图片文件名 (不包含扩展名) 和尺寸
		\caption{The warp factor in Model B  with different values of the parameters $b$ and $\alpha$.}  % 图片说明
		\label{AAAAAAA}  % 图片标签，方便引用
	\end{figure}

	In this model, we consider a warp factor with a plateau, which will result in a rich internal structure of the thick brane.  The form of the warp factor is chosen as
	\begin{align}
		A(y) & = \ln(\tanh \left (ky + b \right) - \tanh \left( ky - b \right)),  
	\end{align}
	where $b$ is a parameter that determines the width of the plateau, as shown in Fig.~\ref{AAAAAAA}.
	
\begin{figure}[ht]  % 图片浮动体环境
	\centering        % 图片居中
	\includegraphics[width=0.3\textwidth]{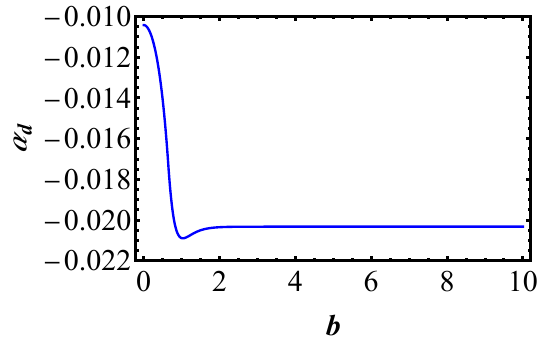}  % 图片文件名 (不包含扩展名) 和尺寸
	\caption{ The relationship between the lower limit $\alpha_d$ of the parameter $\alpha$ and the parameter $b$ when $\varphi'^2 \geqslant 0$.}  % 图片说明
	\label{alphalimit}  % 图片标签，方便引用
\end{figure}

	The equation for the scalar field is  
   \begin{eqnarray}
   	\varphi'^2(y) &= &16 \alpha  A''''(y)+128 \alpha  A''(y)^2-3 A''(y)\notag\\
   	&+&64 \alpha  A'''(y) A'(y)+40 \alpha  A'(y)^2 A''(y)
   	\big). 
   \end{eqnarray}
   $\varphi'^2 \geqslant 0$ implies $\frac{\alpha_d}{k^2}\leqslant\alpha<\frac{1}{160k^2}$, where the lower limit $\alpha_d$ is a function related to $b$, as shown in Fig.~\ref{alphalimit}.
   
\begin{figure}[h!]
 	\centering
 	\begin{subfigure}{0.2\textwidth} % 设置子图的宽度
 		\centering
 		\includegraphics[width=\linewidth]{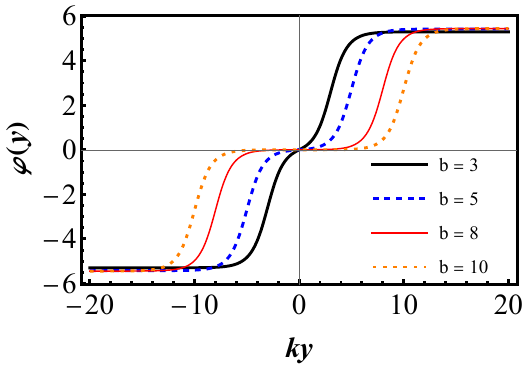} % 替换为你的图片文件名
 		\caption{$\alpha=0$}
 	\end{subfigure}
 	\hspace{0.025\textwidth} % 两张图片之间的水平间距
 	\begin{subfigure}{0.2\textwidth} % 设置子图的宽度
 		\centering
 		\includegraphics[width=\linewidth]{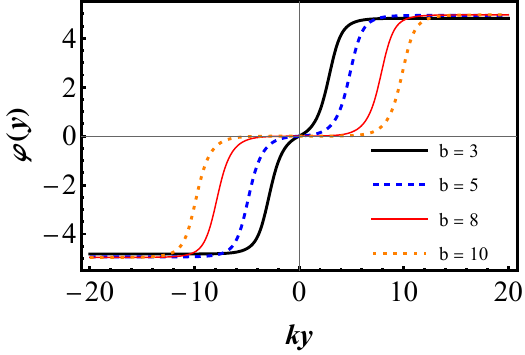} % 替换为你的图片文件名
 		\caption{$\alpha=0.005$}
 		\label{3(c)}
 	\end{subfigure}
 	\vspace{0.5cm} % 两行图片之间的垂直间距
 	\begin{subfigure}{0.2\textwidth} % 设置子图的宽度
 		\centering
 		\includegraphics[width=\linewidth]{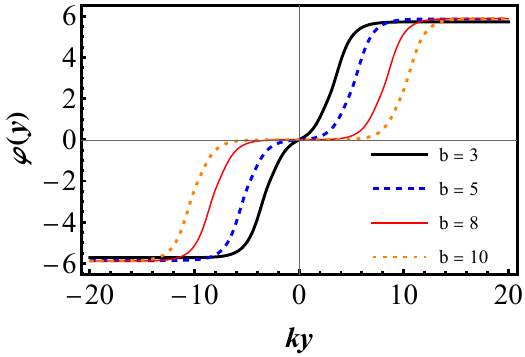} % 替换为你的图片文件名
 		\caption{$\alpha=-0.01$}
 		\label{3(d)}
 	\end{subfigure}
 	\hspace{0.025\textwidth} % 添加水平间距
 	\begin{subfigure}{0.2\textwidth} % 设置子图的宽度
 		\centering
 		\includegraphics[width=\linewidth]{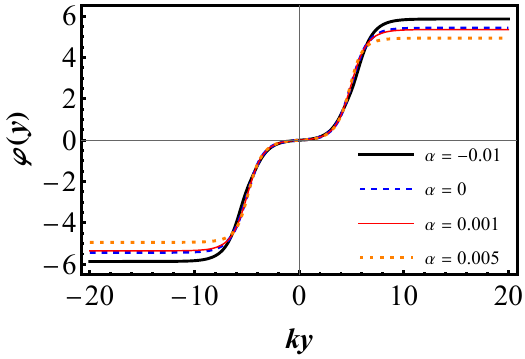} % 替换为你的图片文件名
 		\caption{$b=5$}
 		\label{3(e)}
 	\end{subfigure}
 	\vspace{0.5cm} % 两行图片之间的垂直间距
 	\begin{subfigure}{0.2\textwidth} % 设置子图的宽度
 		\centering
 		\includegraphics[width=\linewidth]{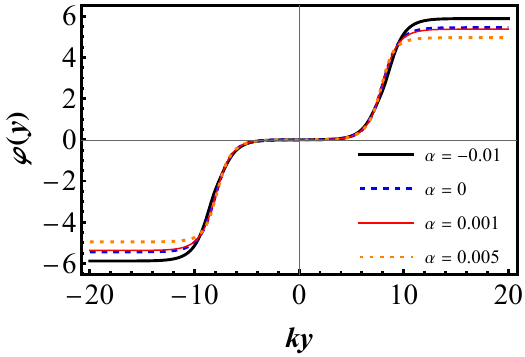} % 替换为你的图片文件名
 		\caption{$b=8$}
 		\label{3(f)}
 	\end{subfigure}
 	\hspace{0.025\textwidth} % 添加水平间距
 	\begin{subfigure}{0.2\textwidth} % 设置子图的宽度
 		\centering
 		\includegraphics[width=\linewidth]{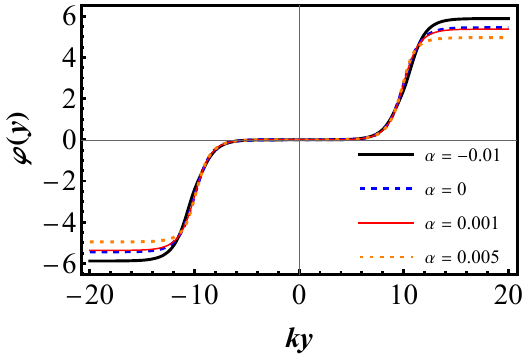} % 替换为你的图片文件名
 		\caption{$b=10$}
 		\label{3(g)}
 	\end{subfigure}
 	\caption{The scalar field $\varphi(y)$ in Model B with different values of the parameters $\alpha$ and $b$.
 	}
 	\label{fig3:two_imagess}
 \end{figure}

\begin{figure}[h!]
		\centering
		\begin{subfigure}{0.2\textwidth} % 设置子图的宽度
			\centering
			\includegraphics[width=\linewidth]{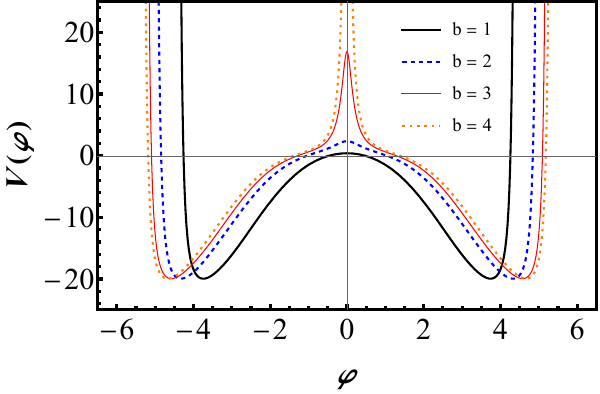} % 替换为你的图片文件名
			\caption{$\alpha=0$}
			\label{3(A)}
		\end{subfigure}
		\hspace{0.025\textwidth} % 添加水平间距
		\begin{subfigure}{0.2\textwidth} % 设置子图的宽度
			\centering
			\includegraphics[width=\linewidth]{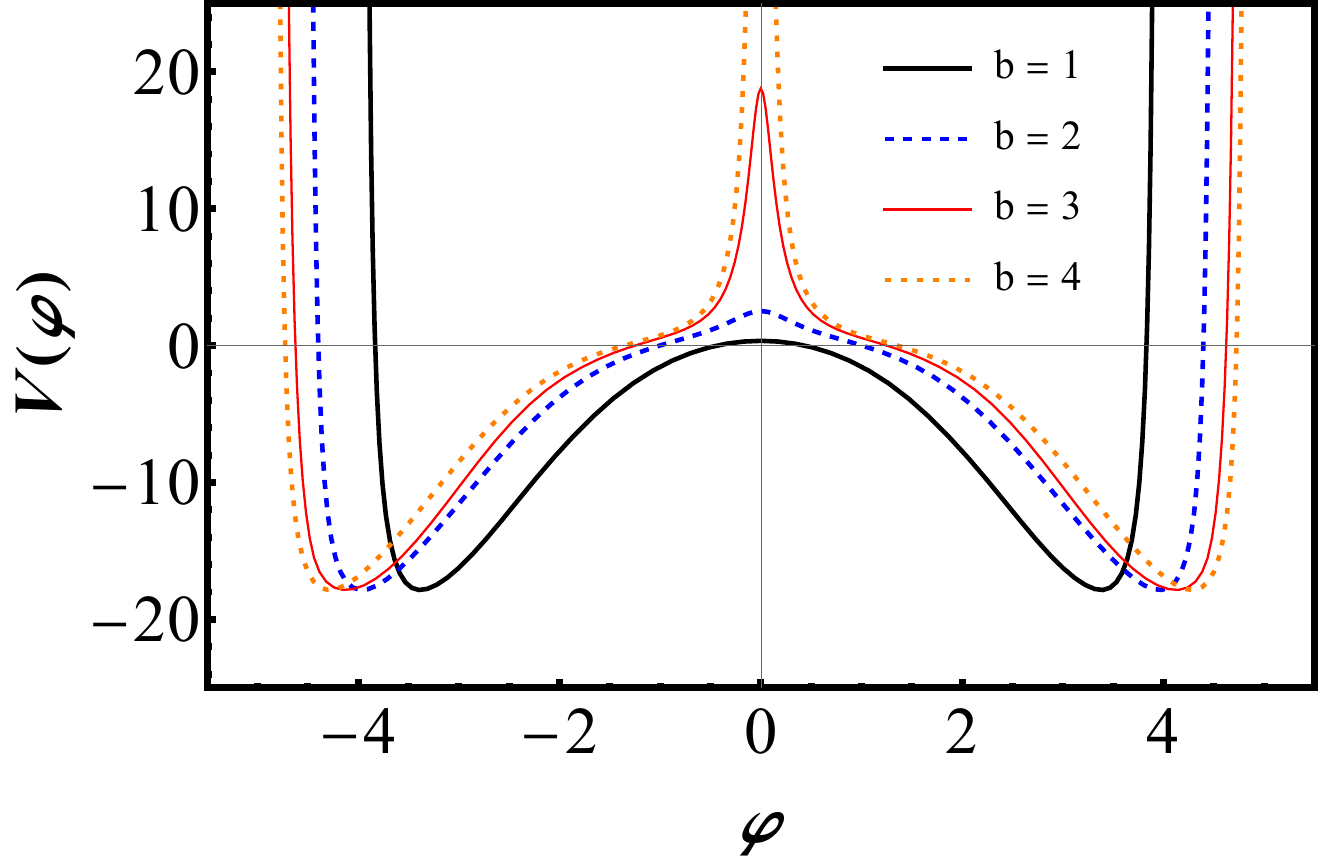} % 替换为你的图片文件名
			\caption{$\alpha=0.005$}
			\label{3(B)}
		\end{subfigure}
		\vspace{0.5cm} % 两行图片之间的垂直间距
		\begin{subfigure}{0.2\textwidth} % 设置子图的宽度
			\centering
			\includegraphics[width=\linewidth]{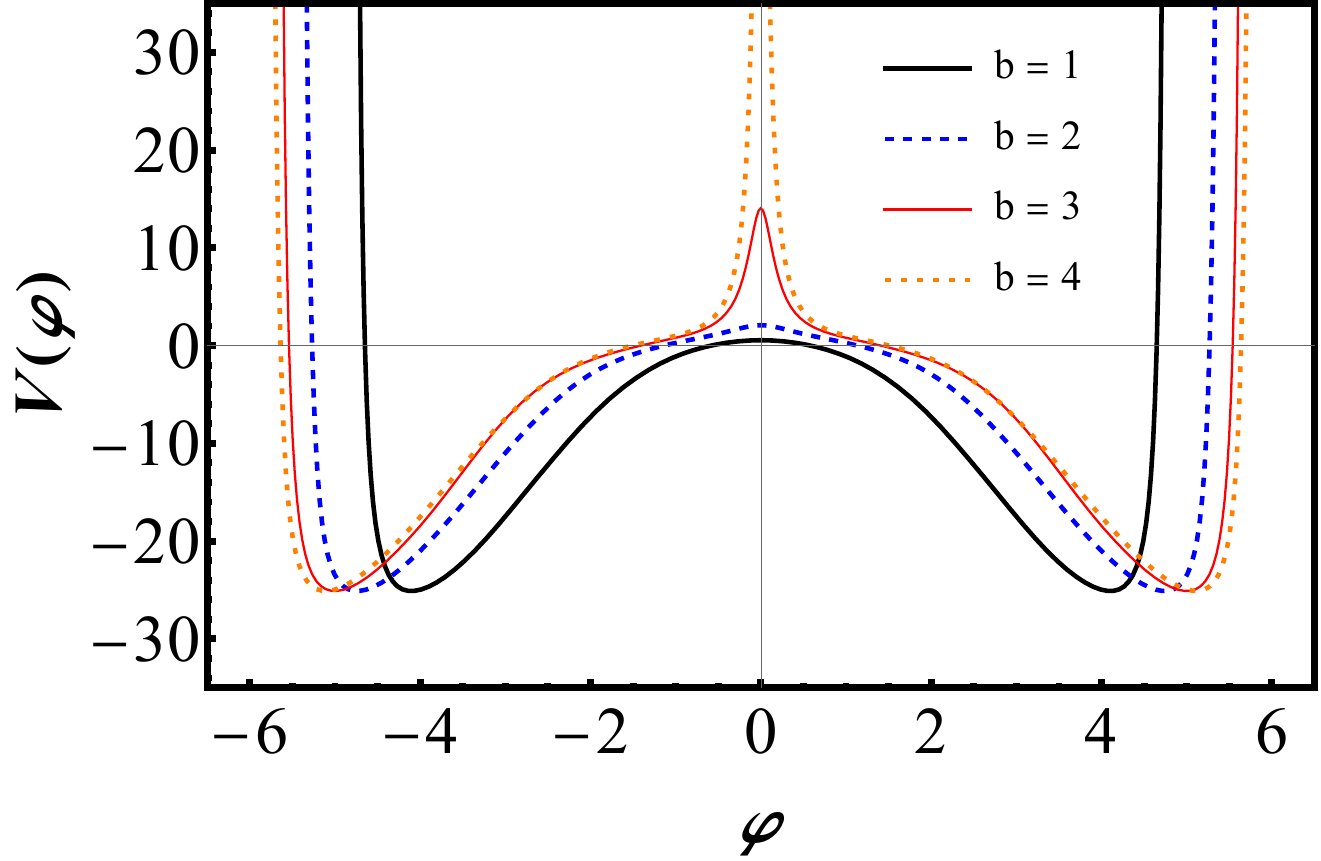} % 替换为你的图片文件名
			\caption{$\alpha=-0.01$}
			\label{3(C)}
		\end{subfigure}
		\hspace{0.025\textwidth} % 添加水平间距
		\begin{subfigure}{0.2\textwidth} % 设置子图的宽度
			\centering
			\includegraphics[width=\linewidth]{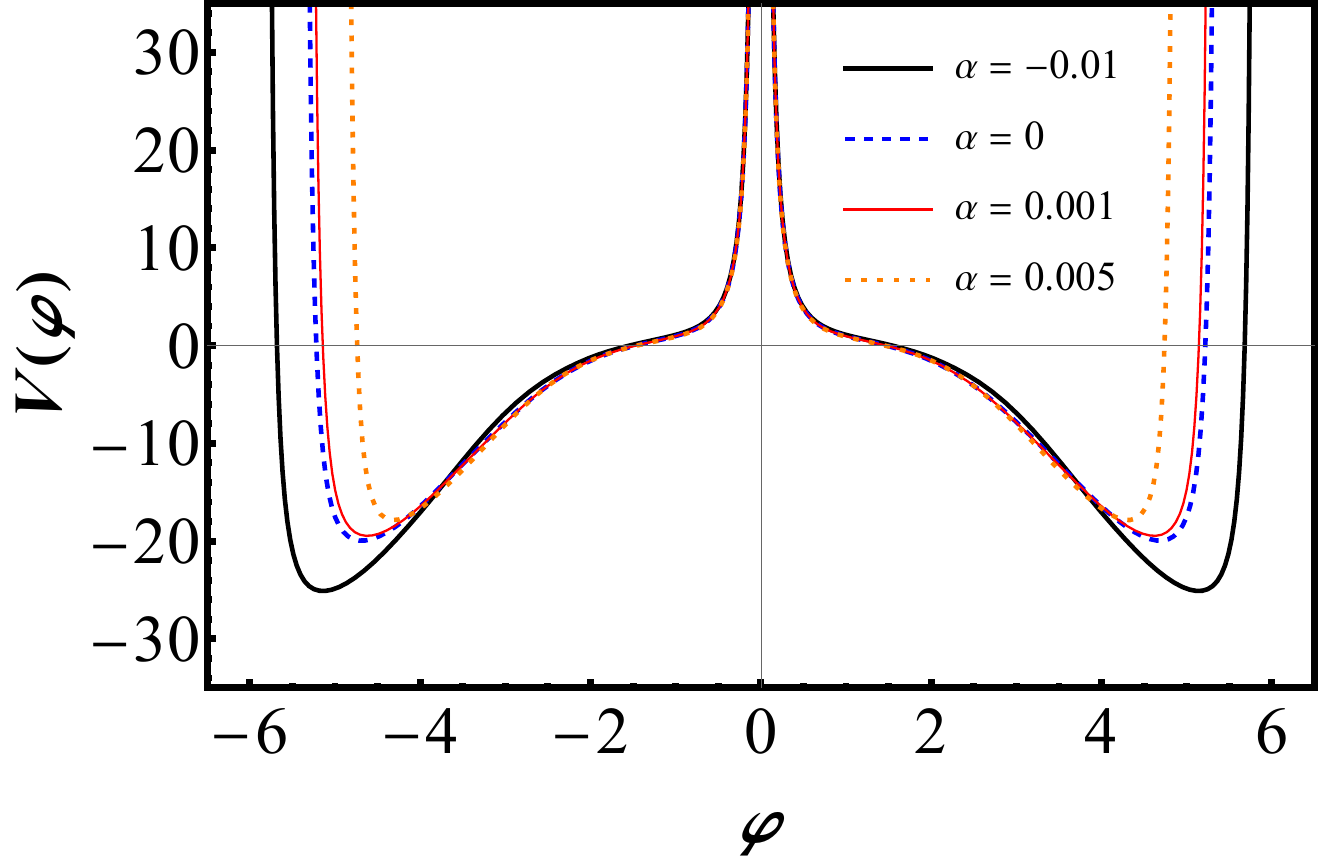} % 替换为你的图片文件名
			\caption{$b = 5$}
			\label{3(D)}
		\end{subfigure}
		\vspace{0.5cm} % 两行图片之间的垂直间距
		\begin{subfigure}{0.2\textwidth} % 设置子图的宽度
			\centering
			\includegraphics[width=\linewidth]{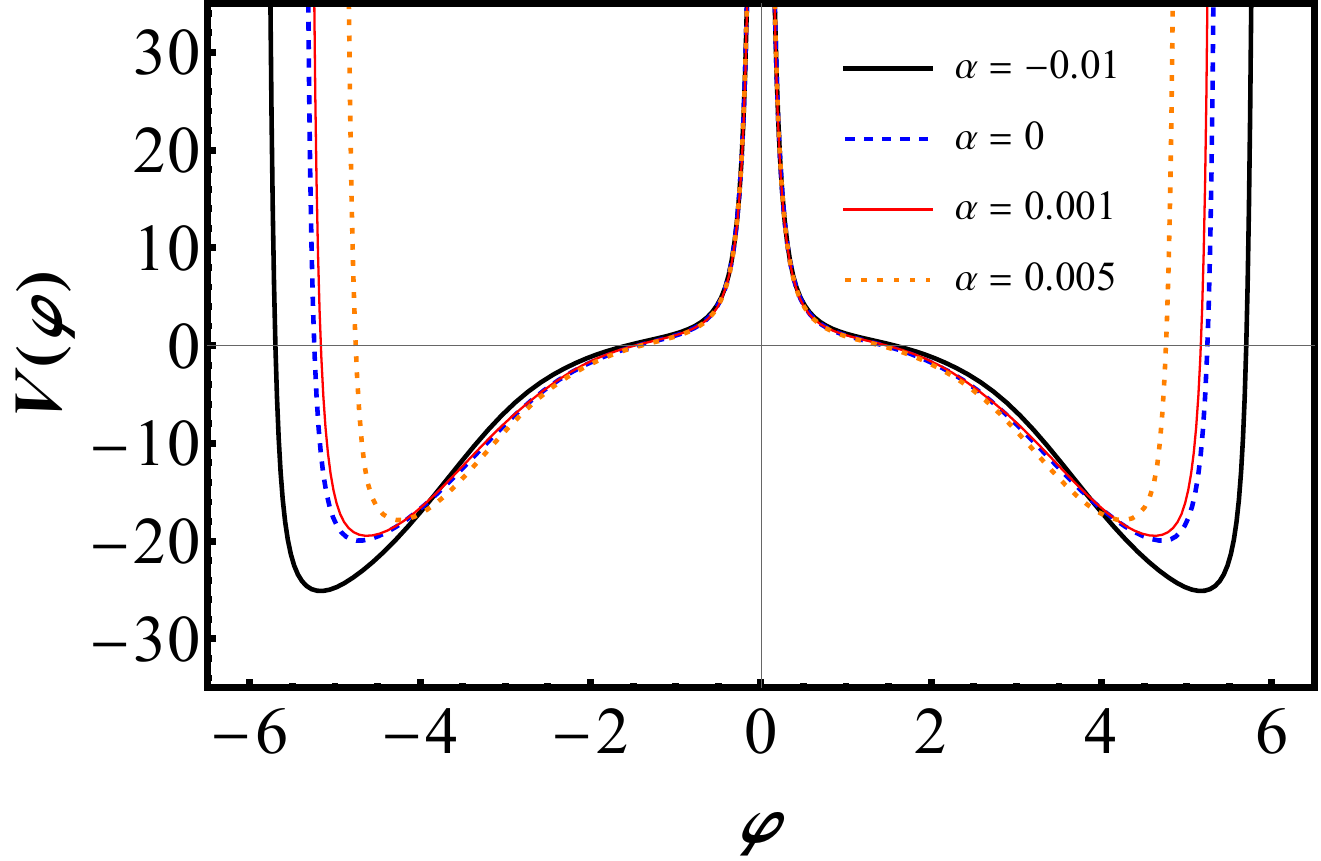} % 替换为你的图片文件名
			\caption{$b = 8$}
			\label{3(E)}
		\end{subfigure}
		\hspace{0.025\textwidth} % 添加水平间距
		\begin{subfigure}{0.2\textwidth} % 设置子图的宽度
			\centering
			\includegraphics[width=\linewidth]{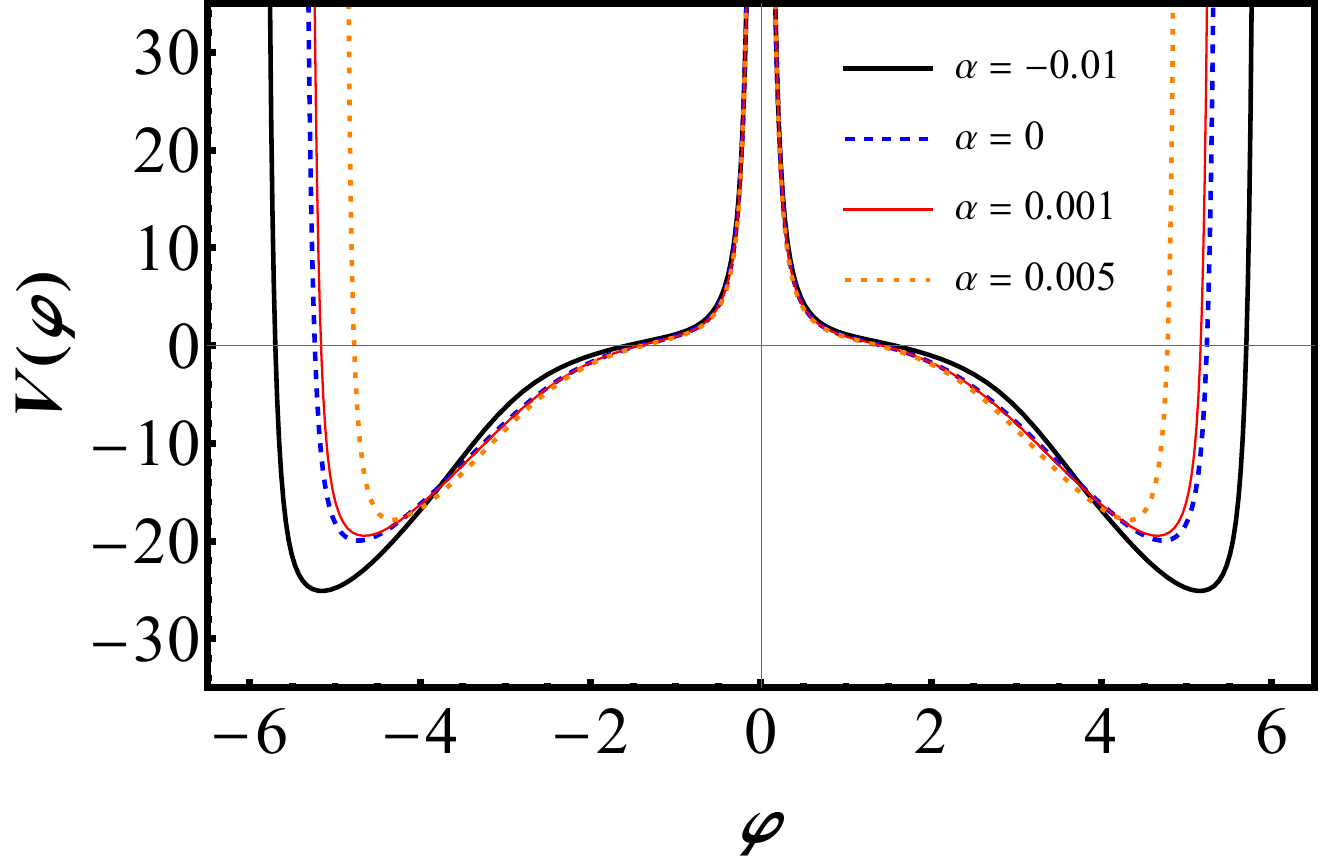} % 替换为你的图片文件名
			\caption{$b = 10$}
			\label{3(F)}
		\end{subfigure}
		\caption{The scalar potential $V(\varphi)$ in Model B with different values of the parameters $b$ and $\alpha$.
		}
		\label{fig8:6_images}
	\end{figure}

 We also consider the $ \mathbb{Z}_2 $ symmetry of the background, the scalar field is chosen to be a kink configuration satisfying $ \varphi(0)=0 $, similar to Model A. In this configuration, the scalar field exhibits a double-kink profile. The parameter $b$ also represents the distance between two sub-kinks and the vacuum expectation value of the scalar field decrease with the parameter $\alpha$, as shown in Fig.~\ref{fig3:two_imagess}. Correspondingly, with fixed $\alpha$, two minima of the scalar potential decrease with $b$ but the value of the scalar potential at $\phi=0$ increases with $b$, as shown in Figs.~\ref{3(A)},~\ref{3(B)}, and~\ref{3(C)}. As shown in Figs.~\ref{3(D)},~\ref{3(E)}, and~\ref{3(F)}, for the case of fixed $b$, two minima and the value at $\varphi=0$ of the scalar potential increase with $b$.

\begin{figure}[ht]  % 图片浮动体环境
	\centering        % 图片居中
	\includegraphics[width=0.3\textwidth]{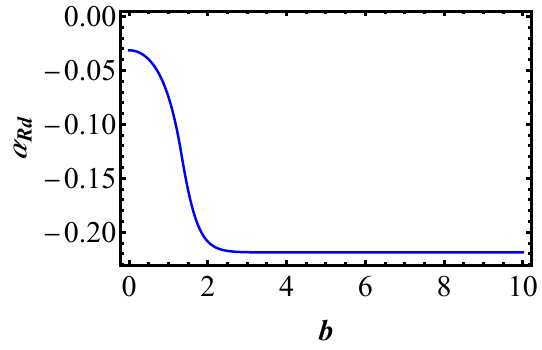}  % 图片文件名 (不包含扩展名) 和尺寸
	\caption{ The relationship between the lower limit $\alpha_{Rd}$ of the parameter $\alpha$ and the parameter $b$ when $f_R > 0$.}  % 图片说明
	\label{alphalimittoRb}  % 图片标签，方便引用
\end{figure}

  Similar to Model A, imposing the stability conditions $f_R > 0$  yields the following constraint on the parameter $\alpha$:
	\begin{equation}
		\frac{\alpha_{Rd}}{k^2} <  \alpha < \frac{0.00625}{k^2},
	\end{equation}
where the lower limit $\alpha_{Rd}$ is also a function related to $b$, as shown in Fig.~\ref{alphalimittoRb}
	
\begin{figure}[ht]
		\centering
		% 第一行的两张图片
		\begin{subfigure}[b]{0.2\textwidth}
			\centering
			\includegraphics[width=\textwidth]{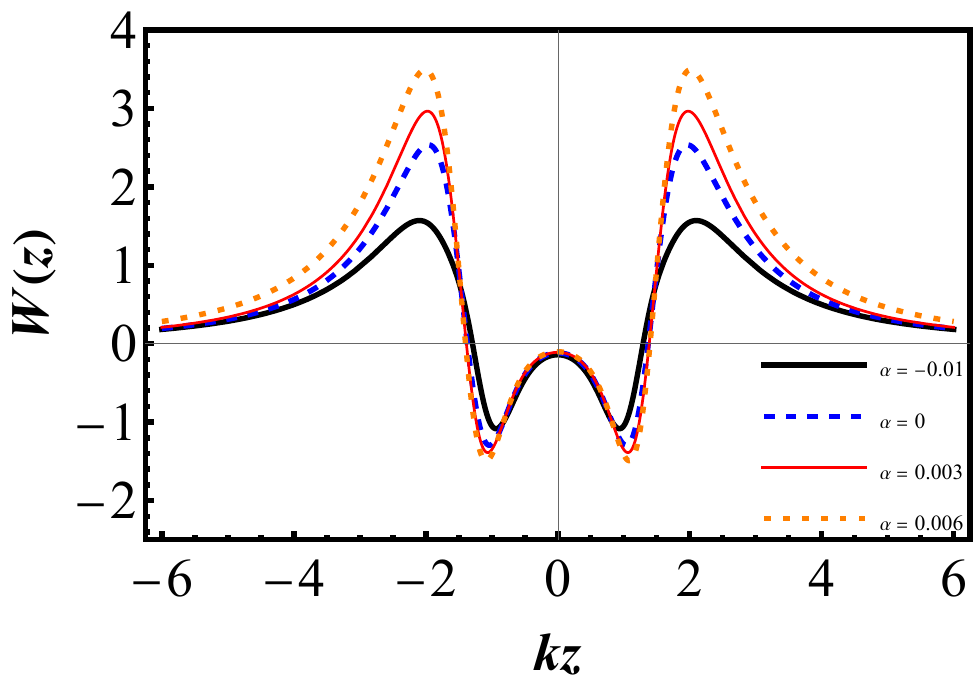} % 第一张图片
			\caption{ $ b = 3 $}
			\label{fig4:img1}
		\end{subfigure}
		\hspace{0.025\textwidth} % 两张图片之间的水平间距
		\begin{subfigure}[b]{0.2\textwidth}
			\centering
			\includegraphics[width=\textwidth]{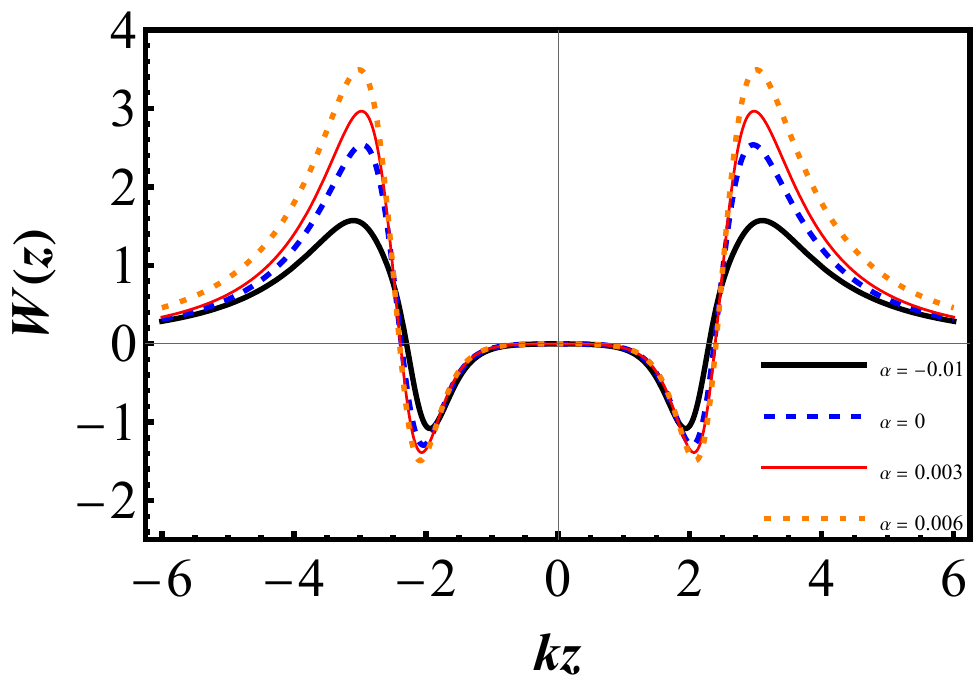} % 第一张图片
			\caption{ $ b = 5 $}
			\label{fig4:img2}
		\end{subfigure}
		\vspace{0.5cm} % 两行图片之间的垂直间距
		\begin{subfigure}[b]{0.2\textwidth}
			\centering
			\includegraphics[width=\textwidth]{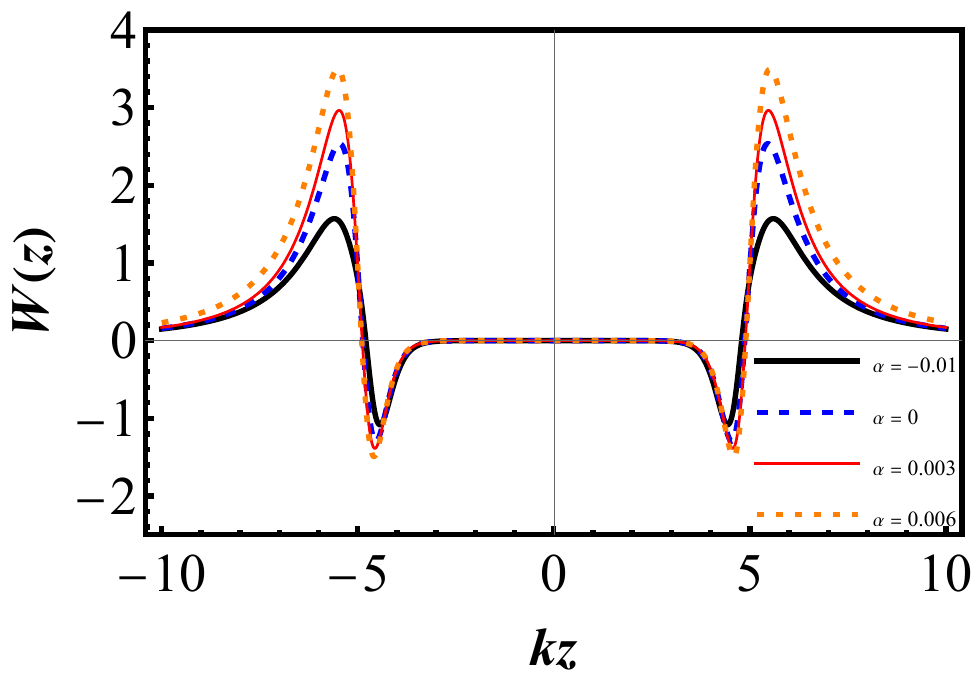} % 第二张图片
			\caption{ $ b = 10 $}
			\label{fig4:img3}
		\end{subfigure}
		\hspace{0.025\textwidth} % 添加水平间距
		\begin{subfigure}{0.2\textwidth} % 设置子图的宽度
			\centering
			\includegraphics[width=\linewidth]{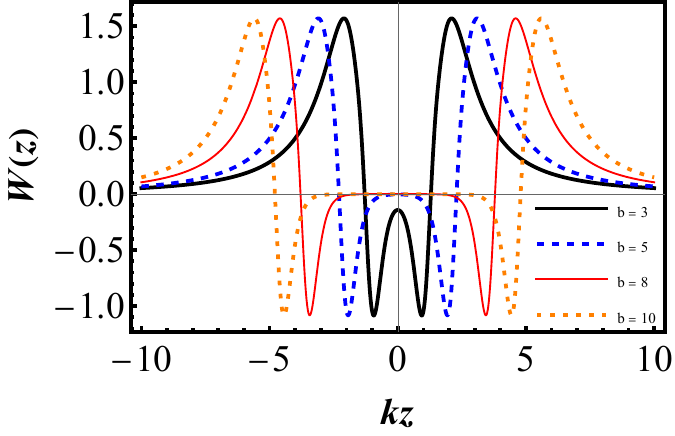} % 替换为你的图片文件名
			\caption{$\alpha=-0.1$}
			\label{fig4:img4}
		\end{subfigure}
		\vspace{0.5cm} % 两行图片之间的垂直间距
		\begin{subfigure}[b]{0.2\textwidth}
			\centering
			\includegraphics[width=\textwidth]{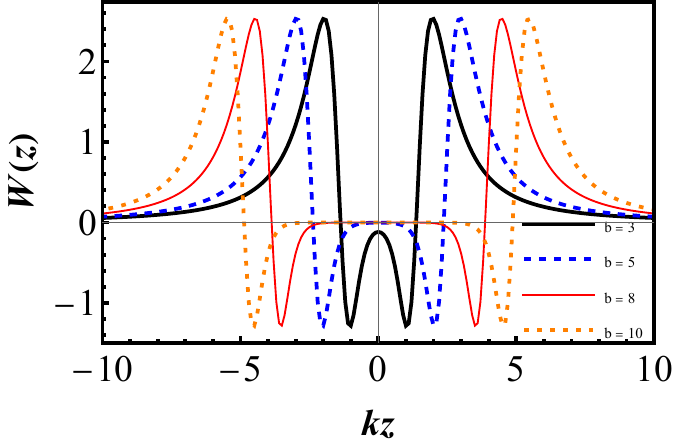} % 第二张图片
			\caption{ $\alpha=0$}
			\label{fig4:img5}
		\end{subfigure}
		\hspace{0.025\textwidth} % 添加水平间距
		\begin{subfigure}{0.2\textwidth} % 设置子图的宽度
			\centering
			\includegraphics[width=\linewidth]{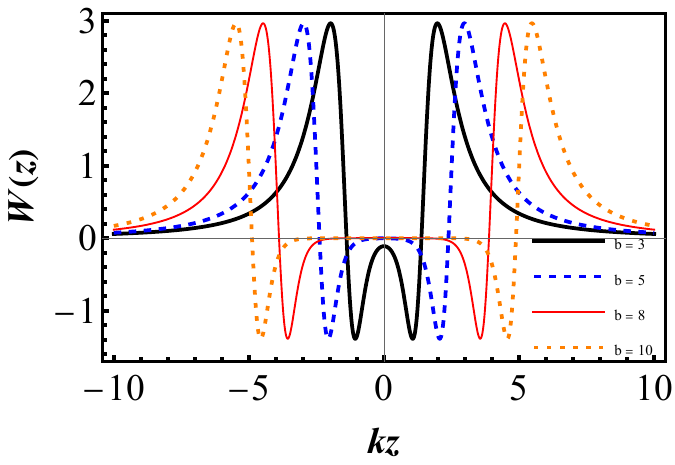} % 替换为你的图片文件名
			\caption{$\alpha=0.003$}
			\label{fig4:img6}
		\end{subfigure}
		\caption{The shapes of the effective potential $ W(z) $ with specific values of $ \alpha$ in Model B.}
		\label{5}
	\end{figure}

Moreover, we present the effective potential $W(z)$  with different values of the parameters $\alpha$ and $b$ in Fig.~\ref{5}. The distance between the two potential barriers increases with the parameter $b$. In contrast, the parameter $\alpha$ plays a decisive role in determining the height of the effective potential barriers. As $\alpha$ increases, the barrier height increases significantly, while variation in barrier width is limited.

	\section{QUASINORMAL MODES OF THICK BRANES}
	\label{sec:main_section}
	In this section, we numerically solve  the Schr\"odinger-like equation  \eqref{13} to obtain the QNMs of the $f(R)$ branes. Three complementary methods are employed: the AIM, the direct integration method, and the time-domain evolution with a Gaussian wave packet. Then, we examine the consistency of the QNM spectra obtained from these independent approaches. Since the effective potential $W(z)$ tends to zero at infinity, the boundary conditions for the Schr\"odinger-like equation \eqref{13} can be set as
	\begin{equation}
		\varphi(z) \propto
		\begin{cases}
			e^{-imkz}, & z \to \infty, \\
			e^{imkz}, & z \to -\infty.
		\end{cases}
		\label{OUT}
	\end{equation}
	
	The AIM proposed by Ciftci et al. \cite{Ciftci2003,Ciftci2005} is a semi-approximate method for solving second-order linear differential equations. In Model A, when $B = 1$ and $\alpha = 0$, the Schrödinger-like equation \eqref{13} can be written as
	\begin{equation}
		-\partial^2_z \psi(z) +  
		\left( \frac{3 \left(-9 + 5 \, \cosh(2 \, \text{arcsinh}(kz)) \right)}{8 (1 + k^2 z^2)^2}-m^2 \right) \psi(z) = 0.
		\label{34}
	\end{equation}
	The AIM requires the first derivative of the equation to be non-zero and is more effective when applied within a finite coordinate range. To ensure that these requirements are satisfied, we introduce a coordinate transformation that maps the infinite domain $z\in(-\infty,+\infty)$ into the finite interval $|u|<1$:
	\begin{equation}
		u = \frac{\sqrt{4k^2 z^2 + 1} - 1}{2kz}.
	\end{equation}
	After performing the above transformation, Eq.~\eqref{34} can be expressed as
	\begin{align}
		&  \left( \frac{3 (-1 + u^2)^2 (2 - 9 u^2 + 2 u^4)}{4 (1 - u^2 + u^4)^2} + m^2 \right) \Psi(u) \nonumber \\
		& + \frac{(-1 + u^2)^3 \left( 2u (3 + u^2) \Psi'(u) + (-1 + u^4) \Psi''(u) \right)}{(1 + u^2)^3}=0.
		\label{38}
	\end{align}
	The boundary conditions~\eqref{OUT} and the function $\psi(u)$ are transformed as follows:
	\begin{equation}
		\psi(u) \sim
		\begin{cases}
			e^{- \frac{im}{k(2u-2)}}, & u \to 1, \\
			e^{\frac{im}{k(2u+2)}}, & u \to -1.
		\end{cases}
	\end{equation}
	Additionally, we express $\psi(u)$ in the form:
	\begin{equation}
		\psi(u) = \xi(u) e^{-\frac{im}{k(2u-2)}} e^{\frac{im}{k(2u+2)}}.
		\label{38a}
	\end{equation}
	Substituting Eq.~\eqref{38a} into Eq.~\eqref{38}, we obtain 
	\begin{equation}
		\xi''(u) = \lambda_0(u) \xi'(u) + s_0(u) \xi(u),
        \label{38qqq}
	\end{equation}
	where
	\begin{align}
		\lambda_0(u) & = -\frac{2 u (-3 + 2 u^2 + u^4 + 2 i (1 + u^2) m)}{(-1 + u^2)^2 (1 + u^2)}, \\
		s_0(u) & = \frac{1}{4 (1 + u^2) (-1 + 2u^2 - 2u^4 + u^6)^2}
		(
		u^{10} (-6 \nonumber \\ 
       &+ 8i m - 4 m^2) + u^4 (57 - 40i m  - 4 m ^2) \nonumber \\
	    &+ u^6 (57 + 40i m  - 4 m ^2) - 2 (3 + 4i m  + 2 m ^2) \nonumber \\
		  &+ u^8 (9 - 24i m  + 4 m ^2) + u^2 (9 + 24i m  + 4 m ^2)
		).
	\end{align}
	We can further derive
	\begin{equation}
		\xi^{\prime\prime\prime}(u) = \lambda_1(u)\xi^{\prime}(u) + s_1(u)\xi(u), 
	\end{equation}
	where
	\begin{align}
		\lambda_1(u) & = \lambda_0^{\prime} + s_0 + \lambda_0^2,  \\
		s_1(u) & = s_0^{\prime} + s_0 \lambda_0. 
	\end{align}
	The AIM uses the recursive structure of Eq.~\eqref{38qqq} to construct a general solution. By continuing to the differentiation process, we obtain the following recursive relations:
	\begin{align}
		\xi_{n+1}(u) & = \lambda_{n-1}(u)\xi^{\prime}(u) + s_{n-1}(u)\xi(u), \\
		\xi_{n+2}(u) & = \lambda_n(u)\xi^{\prime}(u) + s_n(u)\xi(u),
	\end{align}
	where
	\begin{align}
		\lambda_n(u) & = \lambda_{n-1}^{\prime} + s_{n-1} + \lambda_0 \lambda_{n-1}, \\
		s_n(u) & = s_{n-1}^{\prime} + s_0 \lambda_{n-1}. 
	\end{align}
	When $ n \gg 1$ , the AIM introduces an asymptotic form:
	\begin{equation}
		\frac{s_n(u)}{\lambda_n(u)} = \frac{s_{n-1}(u)}{\lambda_{n-1}(u)} = \beta(u),
	\end{equation}
	where $\beta(u)$ is a constant independent of $n$. The QNFs are obtained through the following ``quantization condition'':
	\begin{equation}
		\beta(u) = 0.
	\end{equation}
	
	Next, we introduce the numerical evolution method. We consider a Gaussian wave packet initially localized at $ kz = -30 $. The form of the wave packet is given by
	\begin{align}    
		\Psi(0,z) & = e^{-\frac{(kz + 30)^2}{4}}, \\
		\frac{\partial\Psi(t,z)}{\partial t} \Bigg|_{t=0} & =-\frac{\partial \Psi(t,z)}{\partial z} \Bigg|_{t=0} = \frac{1}{2} (kz + 30) e^{-\frac{(kz + 30)^2}{4}}.
	\end{align}
	We also impose the outgoing boundary condition, similar to Eq.~\eqref{OUT}. The QNFs are obtained  by analyzing the peak frequencies in the Fourier spectrum of the evolved Gaussian wave packet and the  decay rates of the corresponding oscillation modes.
	
\begin{figure}[ht]  % 图片浮动体环境
		\centering        % 图片居中
		\includegraphics[width=0.29\textwidth]{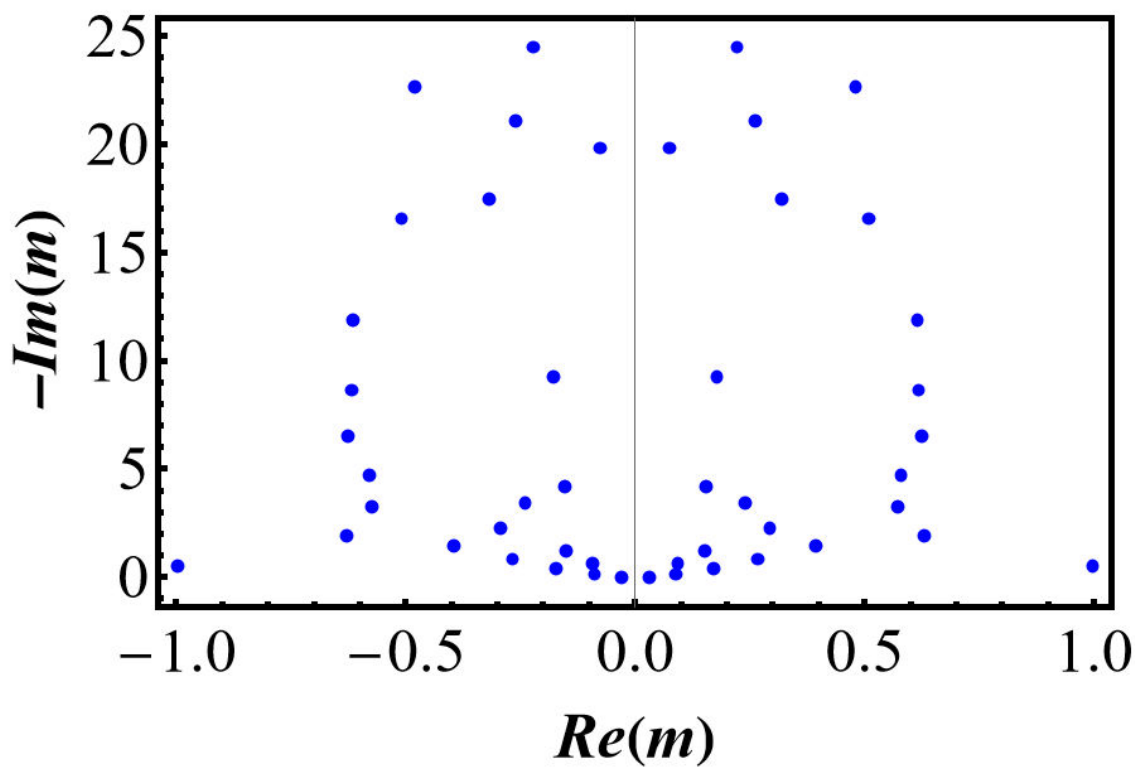}  % 图片文件名 (不包含扩展名) 和尺寸
		\caption{The QNFs of Model A with the AIM ($B = 1$, $\alpha = 0$).}  % 图片说明
		\label{6}  % 图片标签，方便引用
	\end{figure}

	Figure \ref{6} shows the QNFs calculated by the AIM. To validate these results, we need to compare these results with those obtained by the direct integration method and numerical evolution method. The comparison of the results from the three methods is shown in Tab.~\ref{tab-1}. The high consistency among the results obtained by these methods confirms the reliability and applicability of the AIM.
	
	\begin{figure}[h!]
		\centering
		% 第一行的两张图片
		\begin{subfigure}[b]{0.2\textwidth}
			\centering
			\includegraphics[width=\textwidth]{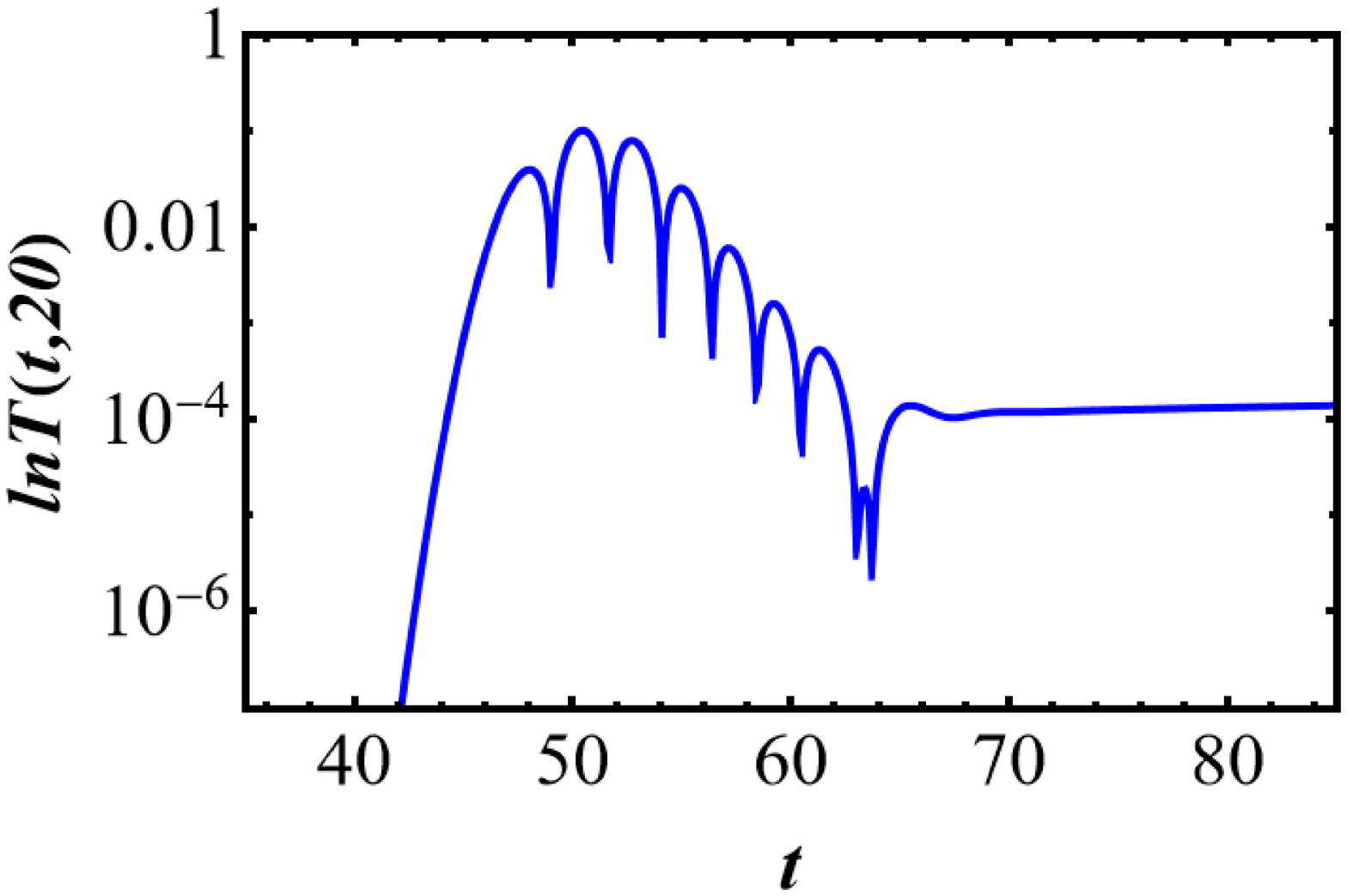} % 第一张图片
			\caption{$B = 1,~\alpha = 0 $}
			\label{7a}
		\end{subfigure}
		\hspace{0.025\textwidth} % 两张图片之间的水平间距
		\begin{subfigure}[b]{0.2\textwidth}
			\centering
			\includegraphics[width=\textwidth]{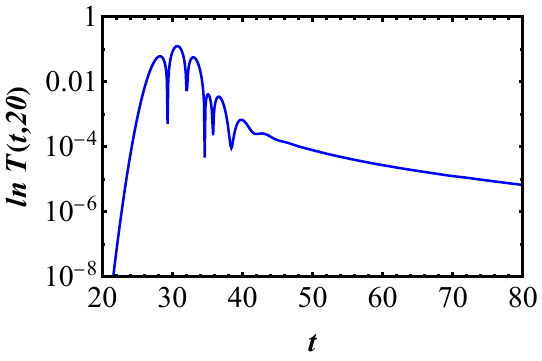} % 第二张图片
			\caption{$B = 3,~\alpha = 0 $}
			\label{7b}
		\end{subfigure}
		
		\vspace{0.5cm} % 两行图片之间的垂直间距
		\begin{subfigure}[b]{0.2\textwidth}
			\centering
			\includegraphics[width=\textwidth]{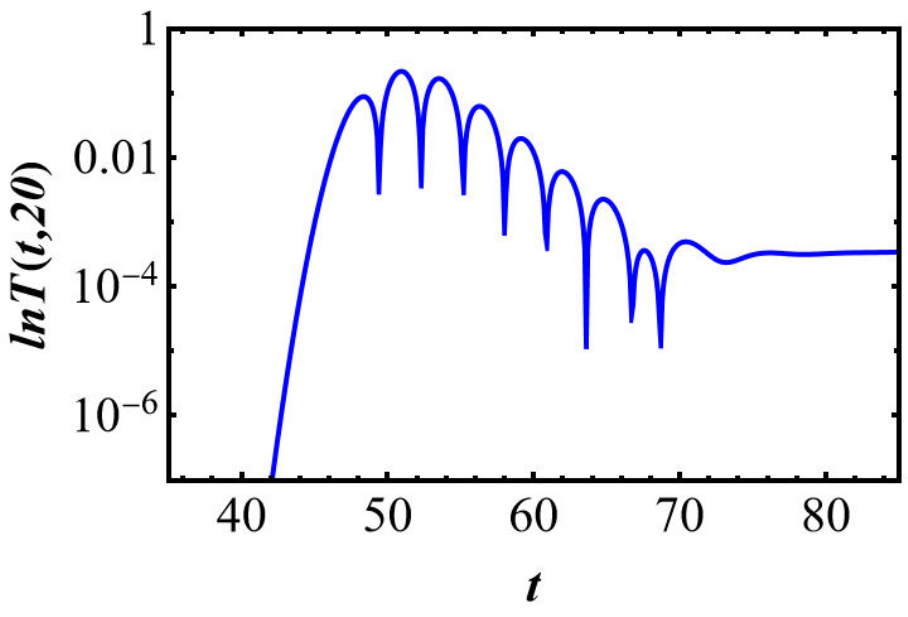} % 第三张图片
			\caption{$B = 1,~\alpha = \alpha 1 $}
			\label{7c}
		\end{subfigure}
		\hspace{0.025\textwidth} % 两张图片之间的水平间距
		\begin{subfigure}[b]{0.2\textwidth}
			\centering
			\includegraphics[width=\textwidth]{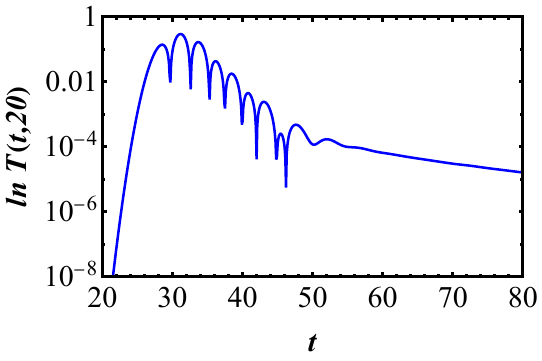} % 第四张图片
			\caption{$B = 3,~\alpha = \alpha 1 $}
			\label{7d}
		\end{subfigure}
		
		\vspace{0.5cm} % 两行图片之间的垂直间距
		\begin{subfigure}[b]{0.2\textwidth}
			\centering
			\includegraphics[width=\textwidth]{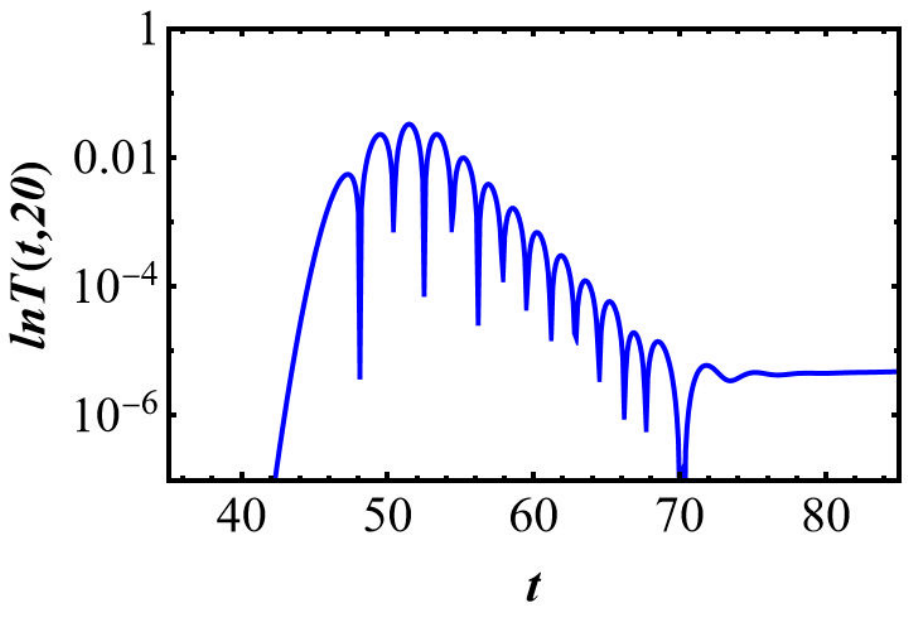} % 第三张图片
			\caption{$B = 1,~\alpha = \alpha 2 $}
			\label{7e}
		\end{subfigure}
		\hspace{0.025\textwidth} % 两张图片之间的水平间距
		\begin{subfigure}[b]{0.2\textwidth}
			\centering
			\includegraphics[width=\textwidth]{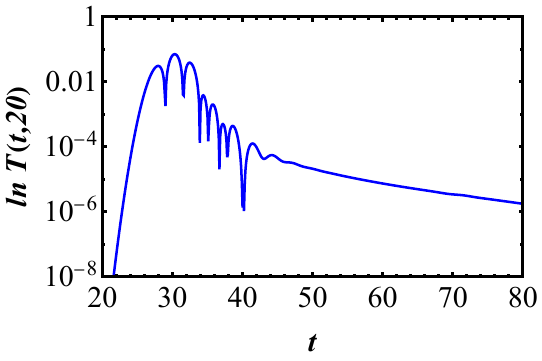} % 第四张图片
			\caption{$B = 3,~\alpha = \alpha 2 $}
			\label{7f}
		\end{subfigure}
		\vspace{0.5cm} % 两行图片之间的垂直间距
		\begin{subfigure}[b]{0.2\textwidth}
			\centering
			\includegraphics[width=\textwidth]{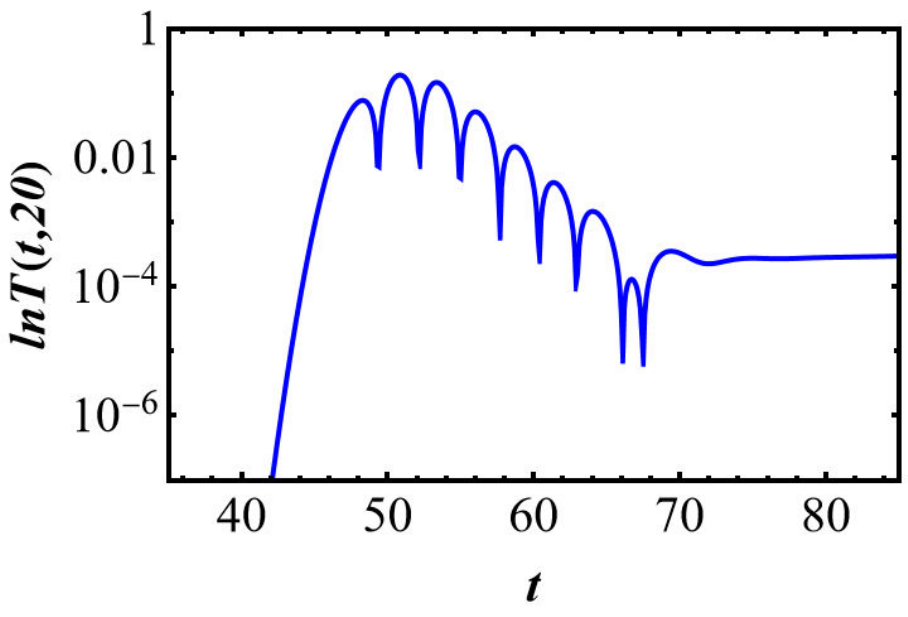} % 第三张图片
			\caption{$B = 1,~\alpha = \alpha s $}
			\label{7g}
		\end{subfigure}
		\hspace{0.025\textwidth} % 两张图片之间的水平间距
		\begin{subfigure}[b]{0.2\textwidth}
			\centering
			\includegraphics[width=\textwidth]{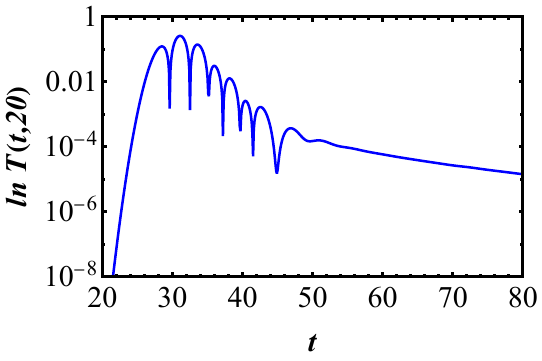} % 第四张图片
			\caption{$B = 3,~\alpha = \alpha s $}
			\label{7h}
		\end{subfigure}
		\caption{The time evolution of the logarithm of the absolute value of the wave function $\ln T(t, z)$ with different values of the parameters $\alpha$ and $B$ at $kz = 20$ in Model A, where $T(t, z) = |H(t, z)|$.}
		\label{7}
	\end{figure}
	
	\begin{figure}[h!]
		\centering
		% 第一行的两张图片
		\begin{subfigure}[b]{0.23\textwidth}
			\centering
			\includegraphics[width=\textwidth]{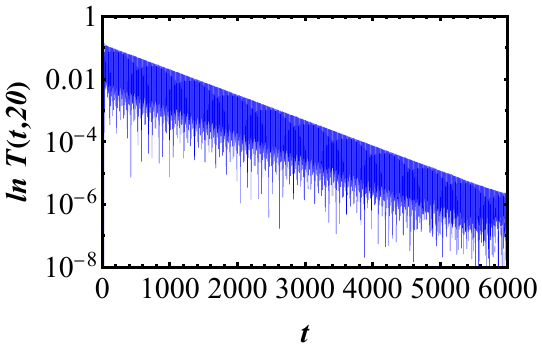} % 第一张图片
			\caption{$b=5,~k^2 \alpha = 0 $}   
			\label{8a}
		\end{subfigure}
		\hspace{0.005\textwidth} % 两张图片之间的水平间距
		\begin{subfigure}[b]{0.23\textwidth}
			\centering
			\includegraphics[width=\textwidth]{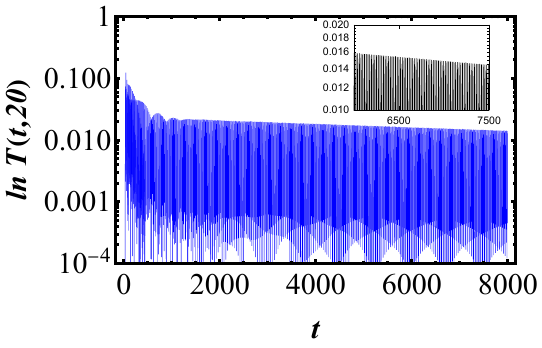} % 第二张图片
			\caption{$b=10,~k^2 \alpha = 0 $}   
			\label{8b}
		\end{subfigure}
		
		\vspace{0.5cm} % 两行图片之间的垂直间距
		\begin{subfigure}[b]{0.22\textwidth}
			\centering
			\includegraphics[width=\textwidth]{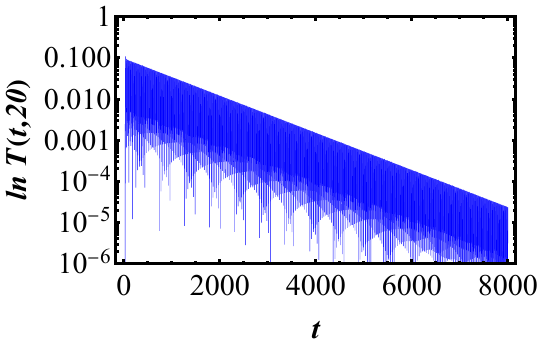} % 第三张图片
			\caption{$b=5,~k^2 \alpha = 0.003 $}
			\label{8c}
		\end{subfigure}
		\hspace{0.005\textwidth} % 两张图片之间的水平间距
		\begin{subfigure}[b]{0.24\textwidth}
			\centering
			\includegraphics[width=\textwidth]{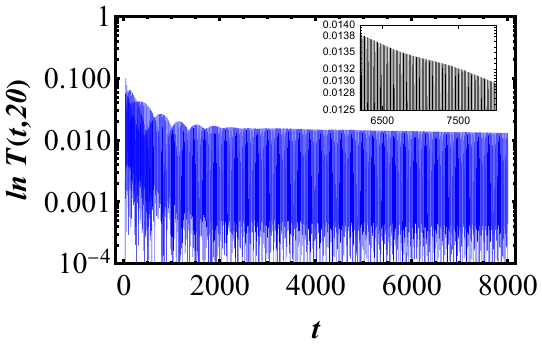} % 第四张图片
			\caption{$b=10,~k^2 \alpha = 0.003 $}
			\label{8d}
		\end{subfigure}
		
		\vspace{0.5cm} % 两行图片之间的垂直间距
		\begin{subfigure}[b]{0.22\textwidth}
			\centering
			\includegraphics[width=\textwidth]{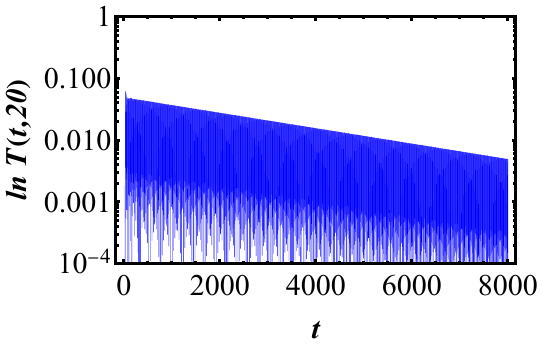} % 第三张图片
			\caption{$b=5,~k^2 \alpha = 0.006 $}
			\label{8e}
		\end{subfigure}
		\hspace{0.005\textwidth} % 两张图片之间的水平间距
		\begin{subfigure}[b]{0.24\textwidth}
			\centering
			\includegraphics[width=\textwidth]{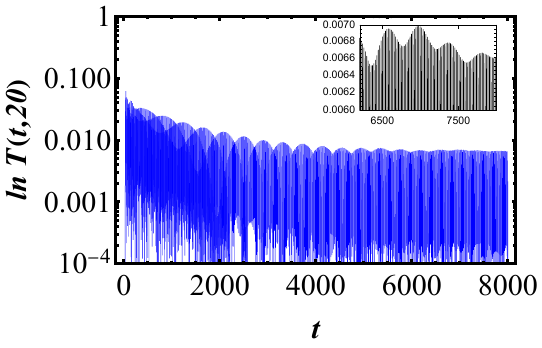} % 第四张图片
			\caption{$b=10,~k^2 \alpha = 0.006 $}
			\label{8f}
		\end{subfigure}
		
		\vspace{0.5cm} % 两行图片之间的垂直间距
		\begin{subfigure}[b]{0.22\textwidth}
			\centering
			\includegraphics[width=\textwidth]{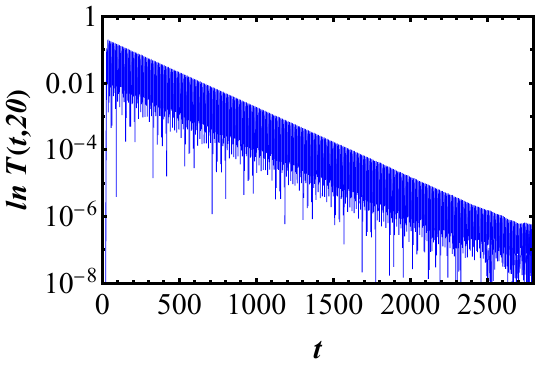} % 第三张图片
			\caption{$b=5,~k^2 \alpha = -0.01 $}
			\label{8g}
		\end{subfigure}
		\hspace{0.005\textwidth} % 两张图片之间的水平间距
		\begin{subfigure}[b]{0.24\textwidth}
			\centering
			\includegraphics[width=\textwidth]{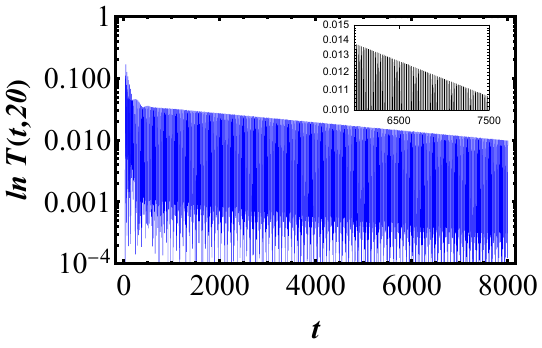} % 第四张图片
			\caption{$b=10,~k^2 \alpha = -0.01 $}
			\label{8h}
		\end{subfigure}
		\caption{The time evolution of $\ln T(t, z)$ with different values of the parameters $\alpha$ and $b$ at $kz = 20$ in Model B, where $T(t, z) = |H(t, z)|$.}
		\label{8}
	\end{figure}
	
	Furthermore, we use the evolution of a Gaussian wave packet to calculate the QNFs. Figure \ref{7}  shows the influence of the parameters $\alpha$ and $B$ on the temporal evolution of a Gaussian wave packet at $kz=20$ in Model A.  As the parameter $\alpha$ increases, the duration of the linear decay stage becomes longer, and the corresponding oscillation frequency also increases. The decay rate and oscillation frequency of the linear decay stage are determined by the corresponding QNMs. As $\alpha$ increases, the real and imaginary parts of the dominant QNF increases, leading to faster decay and higher frequency. In contrast, as the parameter $B$ increases, the duration of the linear decay stage decreases, while the corresponding oscillation frequency increases.
	
	We also investigate the influence of the parameters $\alpha$ and $b$ on the temporal evolution of a Gaussian wave packet at $kz=0$ in Model B, as shown in Fig.~\ref{8}.  As $\alpha$ and $b$ increase, the real part of the dominant QNF decreases, and the decay rate becomes smaller. It is worth noting that, as the value of $\alpha$ increases, the effective potential barrier becomes higher, leading to a reduced damping rate of higher-order modes. As a result, multiple linear decay stages gradually emerge in the evolution waveform. By fitting the data from these linearly decaying waveforms, the real and imaginary parts of the QNFs can be extracted. The fitting results are compared with those obtained by the direct integration method, as shown in Tab. \ref{tab-1}. The maximum relative error in the real part of the QNFs obtained by different methods is about $ 5.04 \% $. In contrast, the deviation in the imaginary part are more significant, particularly in  Model B. This is primarily because the dominant mode exhibits a very small decay rate, which makes numerical fitting more sensitive to errors and thus leads to larger deviations.
	
	\begin{table*}[ht]
		\centering
		\renewcommand{\arraystretch}{1} % 增加行高
		\resizebox{0.75\textwidth}{!}{%
			\begin{tabular}{|c|c|c|c|c|}
				\hline
				\multicolumn{5}{|c|}{Model A} \\
				\hline 
				$\alpha$ & \diagbox{$B$}{Method} & AIM & direct integration & numerical evolution \\
				\hline
				\multirow{4}{*}{$0$} 
				& 1 & $0.997041 - 0.526325\text{ i}$ & $0.997015 - 0.526364\text{ i}$ & $0.992295 - 0.509013\text{ i}$ \\
				& 2 & $1.50896 - 0.630399\text{ i}$ & $1.50875 - 0.629923\text{ i}$ & $1.51407 - 0.600461\text{ i}$ \\
				& 3 & $1.88968 - 0.719737\text{ i}$ & $1.88954 - 0.71998\text{ i}$ & $1.86629 - 0.680936\text{ i}$ \\
				& 4 & $2.20664 - 0.800028\text{ i}$ & $2.2068 - 0.800101\text{ i}$ & $2.24399 - 0.754224\text{ i}$ \\
				\hline
				\multirow{4}{*}{$\alpha_1$} 
				& 1 & $0.790483 - 0.339546\text{ i}$ & $0.790392 - 0.33866\text{ i}$ & $0.791021 - 0.334158\text{ i}$ \\
				& 2 & --- & $1.11662 - 0.40454\text{ i}$ & $1.1088 - 0.390142\text{ i}$ \\
				& 3 & --- & $1.36014 - 0.46546\text{ i}$ & $1.35748 - 0.479002\text{ i}$ \\
				& 4 & --- & $1.56441 - 0.520431\text{ i}$ & $1.51402 - 0.503085\text{ i}$ \\
				\hline
				\multirow{4}{*}{$\alpha_2$} 
				& 1 & $1.30058 - 0.546038\text{ i}$ & $1.3006 - 0.546061\text{ i}$ & $1.29014 - 0.543022\text{ i}$ \\
				& 2 & $1.89566 - 0.513546\text{ i}$ & $1.89562 - 0.5135\text{ i}$ & $1.89253 - 0.528444\text{ i}$ \\
				& 3 & --- & --- & --- \\
				& 4 & --- & --- & --- \\
				\hline
				\multirow{4}{*}{$\alpha_s$} 
				& 1 & $0.843443 - 0.400194\text{ i}$ & $0.842861 - 0.39985\text{ i}$ & $0.845798 - 0.389853\text{ i}$ \\
				& 2 & --- & $1.17727 - 0.459233\text{ i}$ & $1.17443 - 0.43039\text{ i}$ \\
				& 3 & --- & $1.41978 - 0.515852\text{ i}$ & $1.41726 - 0.512214\text{ i}$ \\
				& 4 & --- & $1.62193 - 0.567732\text{ i}$ & $1.59742 - 0.572719\text{ i}$ \\
				\hline
				\multicolumn{5}{|c|}{Model B} \\
				\hline
				$k^2 \alpha$ & \diagbox{$b$}{Method} & \multicolumn{1}{c|}{direct integration} & \multicolumn{2}{c|}{numerical evolution} \\
				\hline
				\multirow{2}{*}{$0$} 
				& 5 & $0.681626 - 0.00186717\text{ i}$ & \multicolumn{2}{c|}{$0.681896 - 0.0018556\text{ i}$} \\
				& 10 & $0.329812 - 0.0000647505\text{ i}$ & \multicolumn{2}{c|}{$0.329307 - 0.000476791\text{ i}$} \\
				\hline
				\multirow{2}{*}{$0.003$} 
				& 5 & $0.684769 - 0.0010436\text{ i}$ & \multicolumn{2}{c|}{$0.685083 - 0.00109792\text{ i}$} \\
				& 10 & $0.330198 - 0.0000343364\text{ i}$ & \multicolumn{2}{c|}{$0.328872 - 0.0000413532\text{ i}$} \\
				\hline
				\multirow{2}{*}{$0.006$} 
				& 5 & $0.687823 - 0.000286823\text{ i}$ & \multicolumn{2}{c|}{$0.687223 - 0.000300845\text{ i}$} \\
				& 10 & $0.330597 - 0.00000583035\text{ i}$ & \multicolumn{2}{c|}{$0.323876 - 0.000007036\text{ i}$} \\
				\hline
				\multirow{2}{*}{$-0.01$} 
				& 5 & $0.670453 - 0.00483726\text{ i}$ & \multicolumn{2}{c|}{$0.676796 - 0.004900\text{ i}$} \\
				& 10 & $0.328368 - 0.000166777\text{ i}$ & \multicolumn{2}{c|}{$0.328872 -0.000175\text{ i}$} \\
				\hline
			\end{tabular}%
		}
		\caption{The QNFs for Models A and B with different methods and different values of $\alpha$, $b$ and $B$.}
		\label{tab-1}
	\end{table*}
	
\begin{figure}[ht]
		\centering
		% 第一行
		\begin{subfigure}[b]{0.2\textwidth}
			\centering
			\includegraphics[width=\textwidth]{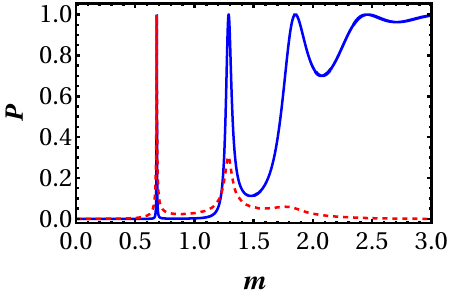}
			\caption{$b=5$, $k^2 \alpha = 0$}
			\label{fig:9a}
		\end{subfigure}
		\hspace{0.025\textwidth} % 两张图片之间的水平间距
		\begin{subfigure}[b]{0.2\textwidth}
			\centering
			\includegraphics[width=\textwidth]{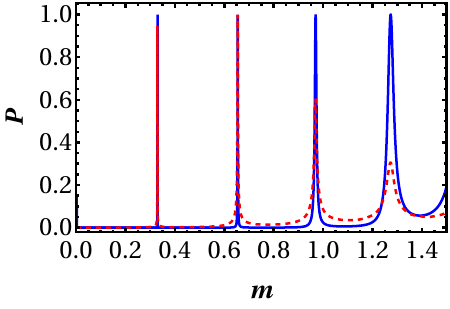}
			\caption{$b=10$, $k^2 \alpha = 0$}
			\label{fig:9d}
		\end{subfigure}
		\vspace{0.5cm} % 两行图片之间的垂直间距
		% 第二行
		\begin{subfigure}[b]{0.2\textwidth}
			\centering
			\includegraphics[width=\textwidth]{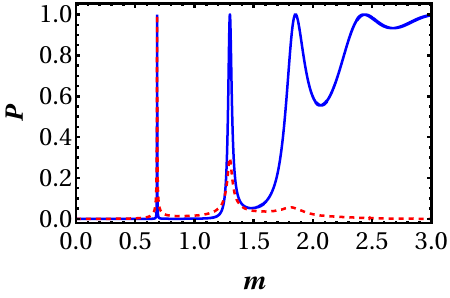}
			\caption{$b=5$, $k^2 \alpha = 0.003$}
			\label{fig:9b}
		\end{subfigure}
		\hspace{0.025\textwidth} % 两张图片之间的水平间距
		\begin{subfigure}[b]{0.2\textwidth}
			\centering
			\includegraphics[width=\textwidth]{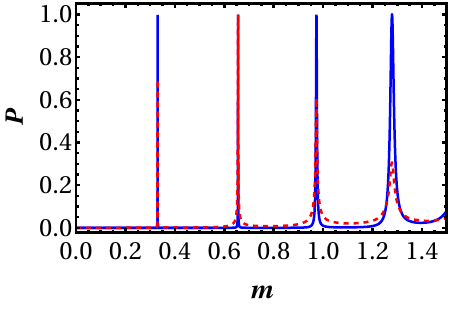}
			\caption{$b=10$, $k^2 \alpha = 0.003$}
			\label{fig:9e}
		\end{subfigure}
		\vspace{0.5cm} % 两行图片之间的垂直间距
		% 第三行
		\begin{subfigure}[b]{0.2\textwidth}
			\centering
			\includegraphics[width=\textwidth]{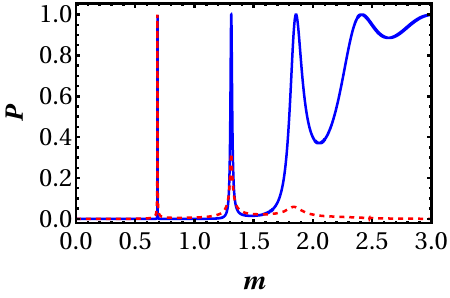}
			\caption{$b=5$, $k^2 \alpha = 0.006$}
			\label{fig:9c}
		\end{subfigure}
		\hspace{0.025\textwidth} % 两张图片之间的水平间距
		\begin{subfigure}[b]{0.2\textwidth}
			\centering
			\includegraphics[width=\textwidth]{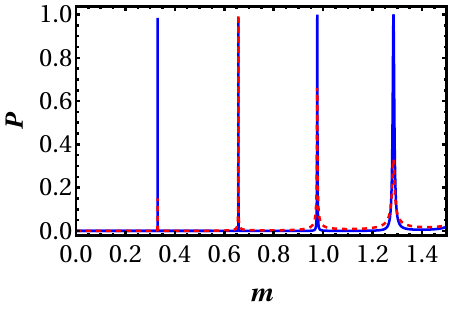}
			\caption{$b=10$, $k^2 \alpha = 0.006$}
			\label{fig:9f}
		\end{subfigure}
		
		\caption{Transmission spectra (blue line) and frequency spectra (red dashed line) for Model B with different values of parameters.}
		\label{fig:six_images}
	\end{figure}

\begin{figure}[ht]
		\centering
		% 第一行的两张图片
		\begin{subfigure}[b]{0.2\textwidth}
			\centering
			\includegraphics[width=\textwidth]{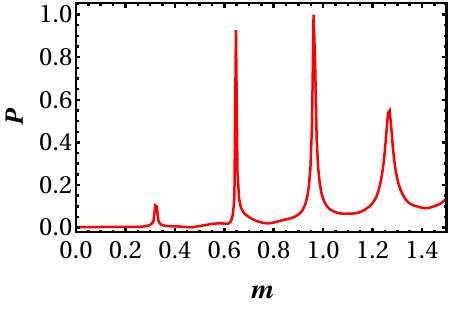} % 第一张图片
			\caption{$0-1000s$}   
			\label{7-1}
		\end{subfigure}
		\hspace{0.025\textwidth} % 两张图片之间的水平间距
		\begin{subfigure}[b]{0.2\textwidth}
			\centering
			\includegraphics[width=\textwidth]{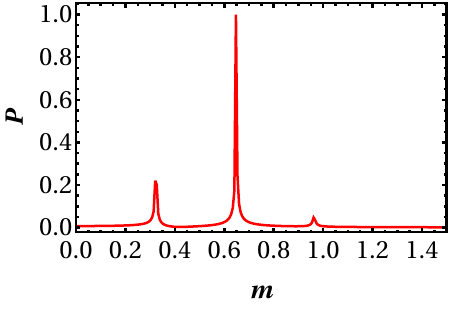} % 第二张图片
			\caption{$1000-2000s$}   
			\label{7-2}
		\end{subfigure}
		
		\vspace{0.5cm} % 两行图片之间的垂直间距
		\begin{subfigure}[b]{0.2\textwidth}
			\centering
			\includegraphics[width=\textwidth]{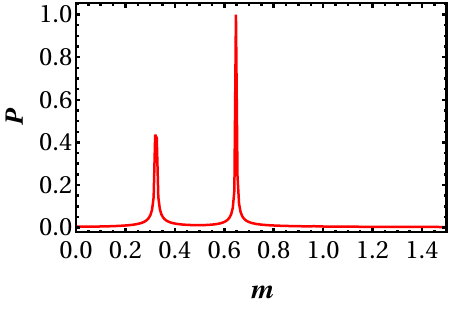} % 第三张图片
			\caption{$2000-3000s$}
			\label{7-3}
		\end{subfigure}
		\hspace{0.025\textwidth} % 两张图片之间的水平间距
		\begin{subfigure}[b]{0.2\textwidth}
			\centering
			\includegraphics[width=\textwidth]{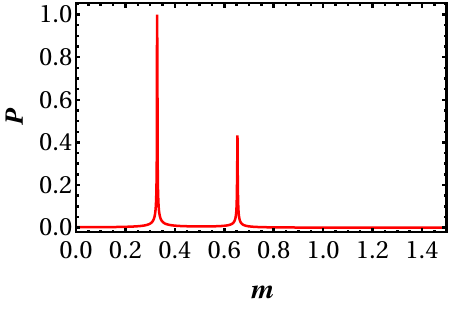} % 第四张图片
			\caption{$3000-8000s$}
			\label{7-4}
		\end{subfigure}
		\caption{Frequency spectra of the wave function evolution at \( kz = 20 \) for Model B with the \( b = 10 \) and \( k^2 \alpha = 0 \).}
		\label{7788}
	\end{figure}

	It is worth noting that, Model B exhibits significantly lower damping frequencies compared with Model A. To further investigate this phenomenon, we employ the direct integration method to calculate more QNFs in Model B, as shown in Tab.~\ref{tab-II}. Furthermore, we calculate the transmission spectrum of the effective potential and compute the frequency spectra of the evolved Gaussian wave packet with \( b = 5 \) and \( b = 10\) at $kz = 20$, as shown in Fig.~\ref{fig:six_images}. We find that the real parts of QNFs in Tab.~\ref{tab-II} coincide with the peak frequencies of both the transmission spectrum and the waveform spectrum, as shown in Tab.~\ref{tab-3}. Moreover, with the parameters $b = 10$ and $\alpha = 0$, we calculate the frequency spectra of the evolved waveforms for different time intervals, as shown in Fig.~\ref{7788}. The frequency components corresponding to \( n = 3 \) and \( n = 4 \) gradually decrease, while the component for \( n = 1 \) increases and eventually surpasses that of \( n = 2 \) as the system evolves. The reason is that the damping frequency increases with the oscillation frequency.
	
	\begin{table*}[ht]
		\centering
		\renewcommand{\arraystretch}{1} %
		\resizebox{0.75\textwidth}{!}{%
			\begin{tabular}{|c|c|c|c|c|}
				\hline
				b & \diagbox{$n$}{$k^2 \alpha$}  & $0$  & $0.003$ & $0.006$  \\ % 以数学模式插入 alpha
				\hline
				\multirow{4}{*}{\centering 5} & 1 & $0.681626 - 0.00186716$ i & $0.684769 - 0.0010436$ i & $0.687824 - 0.000286513$ i  \\
				& 2 & $1.28556 - 0.0283533$ i & $1.29792 - 0.0181109$ i & $1.31031 - 0.00845651$ i \\
				& 3 & $1.81443 - 0.130727 $ i & $1.83002 - 0.0964314$ i &$1.84503 - 0.0625484$ i \\
				& 4 & $2.33816 - 0.320458$ i  & $2.34067 - 0.261878$ i  &$2.33888 - 0.205653$ i\\
				\hline
				\multirow{4}{*}{\centering 10} & 1 & $0.329812 - 0.0000647505$ i & $0.330198 - 0.0000343364$ i & $0.330597 - 0.00000583035$ i  \\
				& 2 & $0.654464 - 0.00078022$ i & $0.655877 - 0.000427545$ i & $0.657199 - 0.000112149$ i  \\
				& 3 & $0.969566-0.00406283$ i & $0.972978 - 0.00233787 $ i& $0.976349 - 0.000823594$ i  \\
				& 4 & $1.27261 - 0.0142411$ i  & $1.27859 - 0.00880127 $ i & $1.28462 - 0.00391664$ i \\
				\hline
			\end{tabular}%
		}
		\caption{The QNFs obtained by the direct integration method for Model B with $ b = 5 $ and $ b = 10 $.} 
		\label{tab-II}
	\end{table*}

	\begin{table*}[ht]
		\centering
		\renewcommand{\arraystretch}{1.5} %  
		\begin{tabular}{|c|c|c|c|c|}
			\hline
			method & $1$  & $2$  & $3$  & $4$  \\ % 以数学模式插入 alpha
			\hline
			transmission spectrum & 0.34641 & 0.648074  & 0.969536  & 1.27279  \\
			\hline
			frequency spectrum & 0.327528 & 0.651907  & 0.966838  & 1.26917   \\
			\hline
			direct integration   & 0.329812 & 0.654464 & 0.969566  &  1.27261  \\
			\hline
		\end{tabular}%
		\caption{The maximum of the transmission spectrum, the maximum of the frequency spectrum of the wave function, and the QNFs obtained by the direct integration method.}
		\label{tab-3}
	\end{table*}
	
	\begin{figure}[ht]
		\centering
		% 第一行的两张图片
		\begin{subfigure}[b]{0.2\textwidth}
			\centering
			\includegraphics[width=\textwidth]{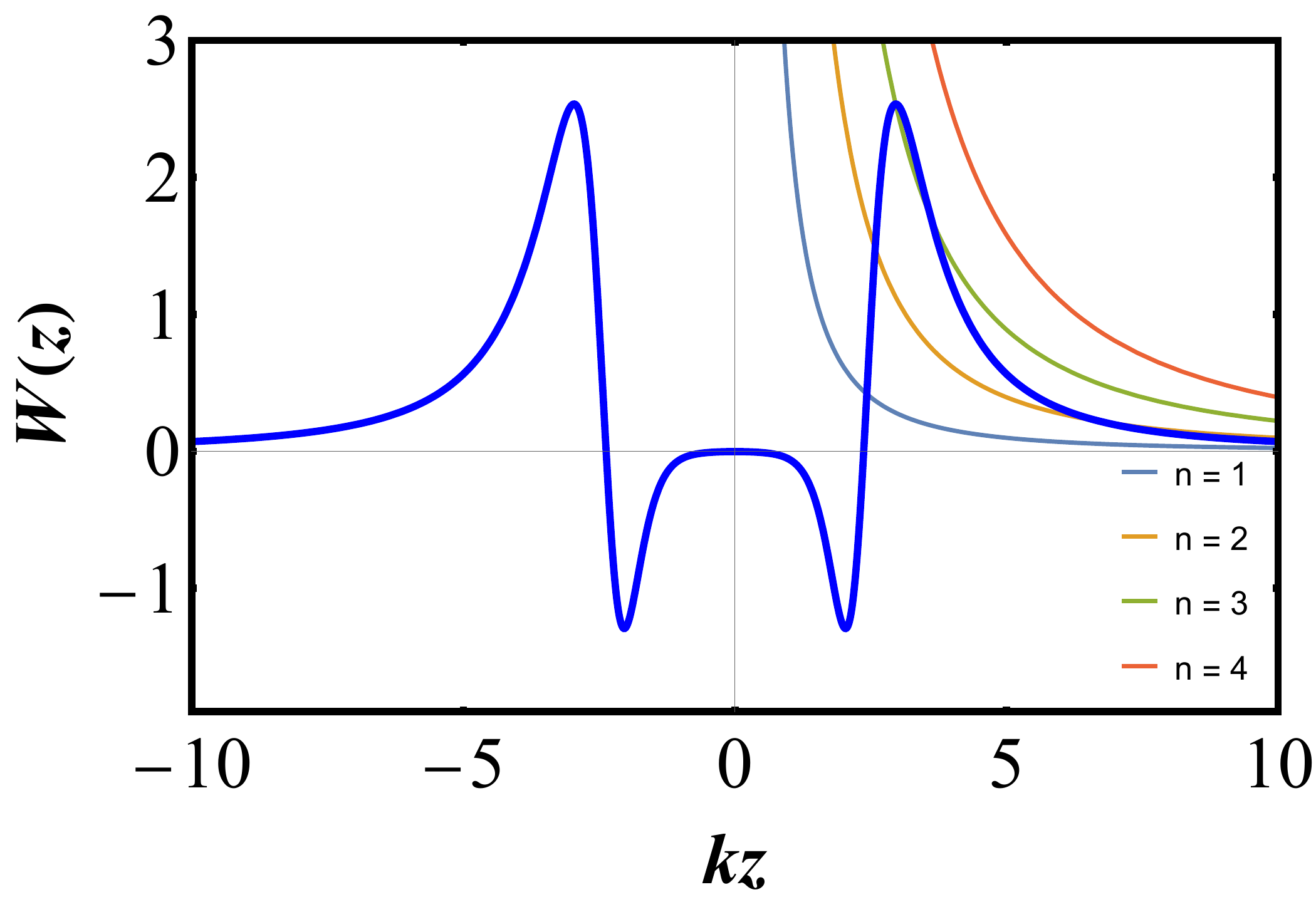} % 第一张图片
			\caption{$b=5, k^2 \alpha = 0 $}
			\label{fig8:img1}
		\end{subfigure}
		\hspace{0.025\textwidth} % 两张图片之间的水平间距
		\begin{subfigure}[b]{0.2\textwidth}
			\centering
			\includegraphics[width=\textwidth]{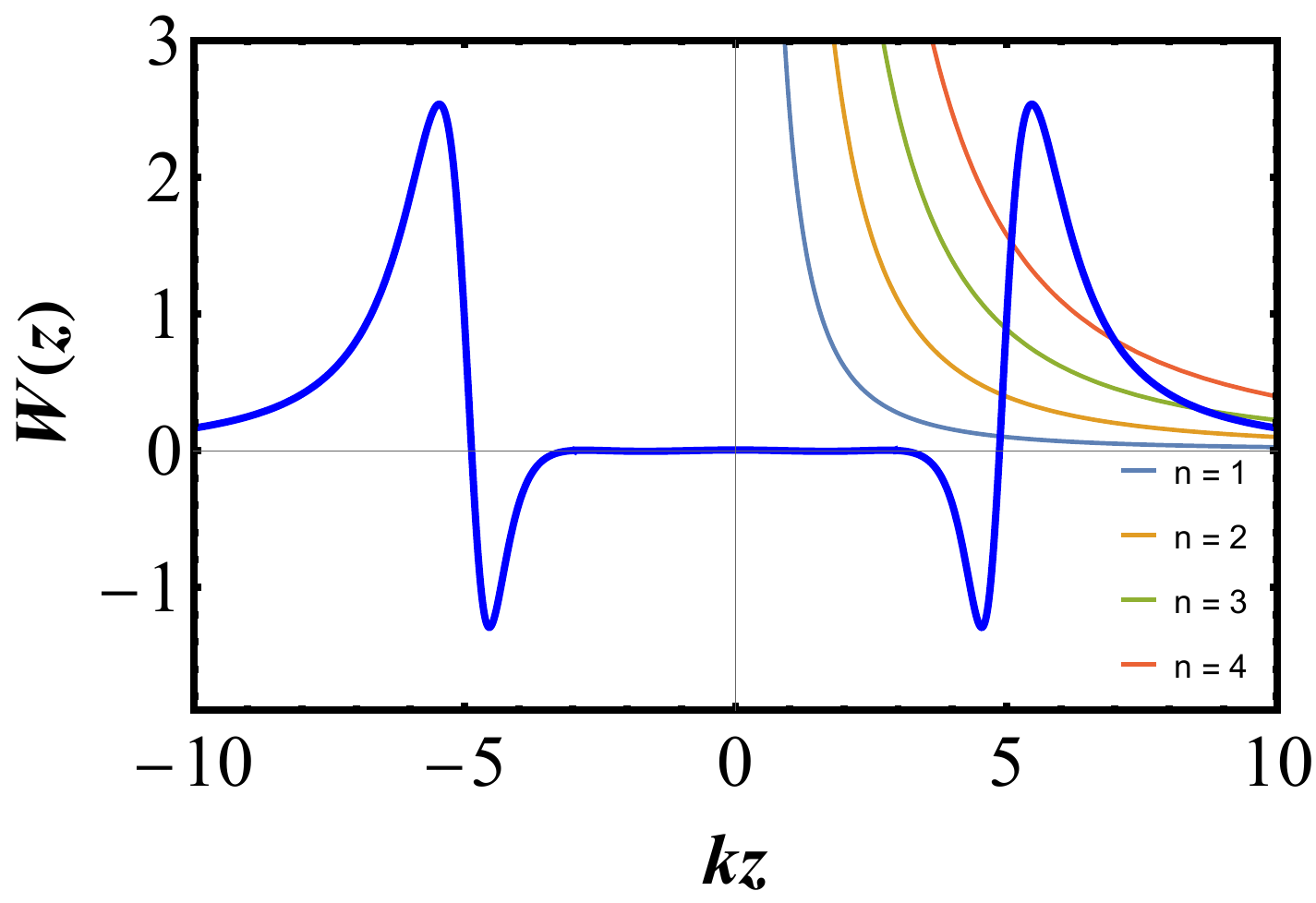} % 第二张图片
			\caption{$b = 10 ,~k^2 \alpha = 0 $}
			\label{fig8:img2}
		\end{subfigure}  
		\caption{Thick blue line: the effective potential for Model B. Thin colored lines: the wavelength–frequency ($\lambda-m^2$) curves for different overtone numbers $n$.}
		\label{fig10}
	\end{figure}
	
	Notably, the QNFs exhibit a nearly arithmetic progression and the real part of the QNFs corresponds to the peaks of the transmission spectrum. The reason is that the wave function forms quasi-localized states between the double barriers. We can approximate the real part of the QNFs by locating the intersection points of the $\lambda - m^2$ curve with the effective potential, as shown in Fig.~\ref{fig10}. When the intersection lies below the barrier height, the accuracy is relatively higher. The results calculated from the intersection points closely match those obtained by the direct integration method, as shown in Tab.~\ref{tab-4}. Moreover, a lower real part of the QNFs corresponds to a higher reflectivity of a single potential barrier, resulting in the wave staying between the barriers for a longer time.
	\begin{table*}
		\centering
		\renewcommand{\arraystretch}{1.5} %  
		\begin{tabular}{|c|c|c|c|c|}
			\hline
			& \multicolumn{2}{c|}{$b=5,~k^2 \alpha=0$} & \multicolumn{2}{c|}{$b=10,~k^2 \alpha=0$} \\
			\hline
			$n$  & $\lambda-m^2$ & direct integration & $\lambda-m^2$ & direct integration \\ 
			\hline
			1 & 0.64807 & 0.681626 & 0.32403  & 0.329812 \\
			\hline
			2 & 1.22065 & 1.28556  & 0.64342  & 0.654464 \\
			\hline
			3 & ---     & 1.81443  & 0.94392  & 0.969566 \\
			\hline
			4 & ---     & 2.33816  & 1.22882  & 1.27261 \\
			\hline
		\end{tabular}%
		\caption{The real part of QNFs calculated by the $\lambda-m^2$ method and the direct integration method for Model B.}
		\label{tab-4}
	\end{table*}
	\section{Conclusion and Discussion}
	\label{sec:conclusion}
	In this article, we investigated the QNMs of thick branes in $f(R)$ gravity. We explored the stability and dynamical behavior of gravitational perturbations in thick brane models. Through the AIM, the direct integration method, and the numerical evolution method, we obtained the QNMs for different background solutions and provided an explanation of their origin from the perspective of the generation mechanism.
	
	Based on the established background solutions, we examined how different parameters affect the effective potential of gravitational perturbations. Furthermore, by studying the potential barrier parameters in Models A and B, we revealed the significant regulatory effects of the barrier steepness and width on the QNFs. Meanwhile, we found that the real parts of the QNFs in Model B exhibit a pattern resembling an arithmetic sequence. The reason is that the quasi-standing waves are formed within the double-barrier structure.

    In braneworld scenarios, the phenomenology of quasinormal modes differs from that of four-dimensional compact objects. The relevant discussions have already been presented in the literature~\cite{Zhu:2024gvl}. In particular, the gravitational echoes predicted in extra-dimensional models do not originate from the properties of the compact object itself, but rather from reflections of gravitational waves off an effective potential barrier in the extra-dimensional direction along the propagation path. As a consequence, unlike the QNMs of compact objects in four-dimensional spacetimes, the echo spectra in braneworld models are largely universal and do not depend sensitively on the parameters of the source, such as its mass or angular momentum. Therefore, if their initial frequency ranges are similar, different gravitational-wave events are expected to share the same resonance spectrum. Correspondingly, the echo frequencies and damping rates are expected to be nearly identical across different events. Furthermore, in extra-dimensional models, gravitational-wave echoes correspond to massive Kaluza-Klein modes. If such modes can be detected on the brane, they would exhibit an effective mass and propagate at speeds slightly lower than the speed of light.
	
	This study provides an important insight into the resonant behavior of gravitational perturbations in thick brane scenarios. We expect that future work could further explore the effects of higher-order curvature corrections on QNFs, thereby advancing the development and application of gravitational theories in more complex spacetimes.

	\section*{Acknowledgments}
	We are thankful to Hai-Long Jia for useful discussions. This work was supported by the National Natural Science Foundation of China (Grants No.~12475056, No.~12247101), the 111 Project under (Grant No.~B20063), the Natural Science Foundation of Gansu Province (No.~22JR5RA389, No.~25JRRA799), and Gansu Province's Top Leading Talent Support Plan.

	\bibliographystyle{unsrt} % 或者你想用的样式
	\bibliography{ref}

\end{document}